\documentclass[twocolumn,floatfix,showpacs,pra,aps]{revtex4}

\usepackage{lastpage}
\usepackage{epsfig}
\usepackage{color}

\newcommand{\ket}[1]{|#1 \rangle}

\begin{document}

\title{Deterministic optical quantum computer using photonic modules}

\author{Ashley M. Stephens$^{1,}{}\footnote{electronic address: a.stephens@physics.unimelb.edu.au}$, Zachary W. E. Evans$^{1}$, Simon J. Devitt$^{2}$, Andrew D. Greentree$^{1}$, Austin G. Fowler$^{3}$, William J. Munro$^{2,4}$, Jeremy L. O'Brien$^{5}$, Kae Nemoto$^{2}$, and Lloyd C. L. Hollenberg$^{1}$}

\affiliation{$^{1}$Centre for Quantum Computer Technology, The University of Melbourne, Victoria 3010, Australia}
\affiliation{$^{2}$National Institute for Informatics, 2-1-2 Hitotsubashi, Chiyoda-ku, Tokyo 101-8430, Japan}
\affiliation{$^{3}$Institute for Quantum Computing, University of Waterloo, Ontario N2L 3G1, Canada}
\affiliation{$^{4}$Hewlett-Packard Laboratories, Filton Road, Stoke Gifford, Bristol BS34 8QZ, United Kingdom}
\affiliation{$^{5}$Centre for Quantum Photonics, H. H. Wills Physics Laboratory \& Department of Electrical and Electronic Engineering, University of Bristol, Merchant Venturers Building, Woodland Road, Bristol, BS8 1UB, United Kingdom}

\date{\today}

\begin{abstract}
The optical quantum computer is one of the few experimental systems to have demonstrated small scale quantum information processing. Making use of cavity quantum electrodynamics approaches to operator measurements, we detail an optical network for the deterministic preparation of arbitrarily large two-dimensional cluster states. We show that this network can form the basis of a large scale deterministic optical quantum computer that can be fabricated entirely on chip.
\end{abstract}


\maketitle

\section{Introduction}
Quantum information science offers a new paradigm for computing. Although there have been many demonstrations of quantum two-level systems, building a large scale quantum computer requires solving problems of scalability, networking, and defect tolerance so that fault tolerant quantum computation can be achieved. Architectures that address some or all of these criteria have been proposed for a number of physical systems, including trapped ions \cite{Kielpinski1}, solid state systems \cite{Taylor1, Hollenberg1}, and superconducting systems \cite{Fowler2}.

In recent years, optical systems \cite{Knill1, Kok1} have emerged as one of the most promising platforms for quantum information processing. In some ways photons constitute almost ideal qubits because of their well defined Hilbert space, immunity from decoherence, and natural mobility. These advantages are reflected by the rapid experimental progress of optical systems - optical systems have demonstrated control of photonic qubits, quantum gates, and even small quantum algorithms \cite{O'Brien2, Walther1, Kok1}.

Techniques for achieving coupling in optical systems can be divided into two broad categories. The first is exemplified by non-linear optical gates such as the optical Fredkin gate \cite{Milburn1} and weak non-linear interactions \cite{Munro1}. The second involves the use of linear elements, photonic measurement, and post-selection. While non-linear techniques remain largely theoretical, linear techniques for optical coupling have shown early experimental success \cite{O'Brien2} and more advanced techniques have been proposed and demonstrated \cite{Kok1}. Such techniques are inherently non-deteministic and so place constraints on the scalability of the optical quantum computer.

The cluster state model of computation \cite{Raussendorf1} is particularly suited to optical systems \cite{Nielsen1} partly because it reduces some of the overheads associated with non-deterministic coupling. Non-deterministic gates can be used to grow a sufficiently large cluster state until the growth rate exceeds some critical rate \cite{Kieling2}, after which it can begin to be consumed by measurement to perform computation. However, non-determinism still limits the operation of the computer as ancillary qubits, conditional routing, and quantum memory are necessary. 

Atom-cavity systems provide effective photon-photon interactions that can be used to achieve deterministic coupling in optical systems. This is exemplified by the photonic module \cite{Devitt1}, which generates a native operator measurement across multiple qubits. The module comprises a single atomic system placed in a high quality cavity and is entirely deterministic in its operation and action. Though it was initially proposed as a simple device to create Bell and Greenberger-Horne-Zeilinger states for quantum communication, quantum key distribution, secret sharing, and dense coding, it was also demonstrated that the module had the flexibility to prepare any stabilizer state and hence could prepare cluster states. Here we explicitly show how to efficiently prepare arbitrarily large two-dimensional cluster states using a parallel network of photonic modules. This modular network may form the basis of a scalable optical quantum computer.

The cluster preparation network is formed by a classically connected pattern of identical devices, each of which may be independently fabricated and characterized before insertion into the network to ensure tolerance against defects. Because of the determinism of the photonic module, the cluster is generated continuously from unentangled single photons. The detector network required to perform computation can therefore be placed immediately after the preparation network so that the cluster is consumed as it is created. This is expected to reduce susceptibility to decoherence and eliminates any need for quantum memory.

Importantly, recent results suggest that our proposal is realizable using existing and near-term technology. Cavity quantum electrodynamics has been shown to induce a single photon Kerr nonlinearity \cite{Turchette1} and has formed the basis of many proposals for quantum gates - for example, Refs.~\cite{Domokos1, Duan1, Duan2}. These proposals have begun to be realized \cite{Rauschenbeutel1} and strong coupling between semiconductor quantum dots and photonic crystal cavities has been reported \cite{Yoshie1, Hennessy1, Englund1}. One of the defining experiments to demonstrate non-linear interactions - photonic blockade - has recently been carried out \cite{Birnbaum1, Dayan1}. State of the art cavity structures include those realized with individual trapped atoms  \cite{Boozer1, Trupke1} and high-Q solid state cavities are being fabricated \cite{Song1}. Finally, one remarkable alternative is coupling superconducting qubits using a cavity bus \cite{Schuster1}.

This paper is organized as follows: Section \ref{TPM} reviews the internal construction and action of the photonic module, Section \ref{CSP} specifies the cluster preparation networks, and Section \ref{EC} addresses issues related to continuously consuming the cluster state.   

\section{Photonic Module}
\label{TPM}

\begin{figure}
\begin{center}
\resizebox{85mm}{!}{\includegraphics{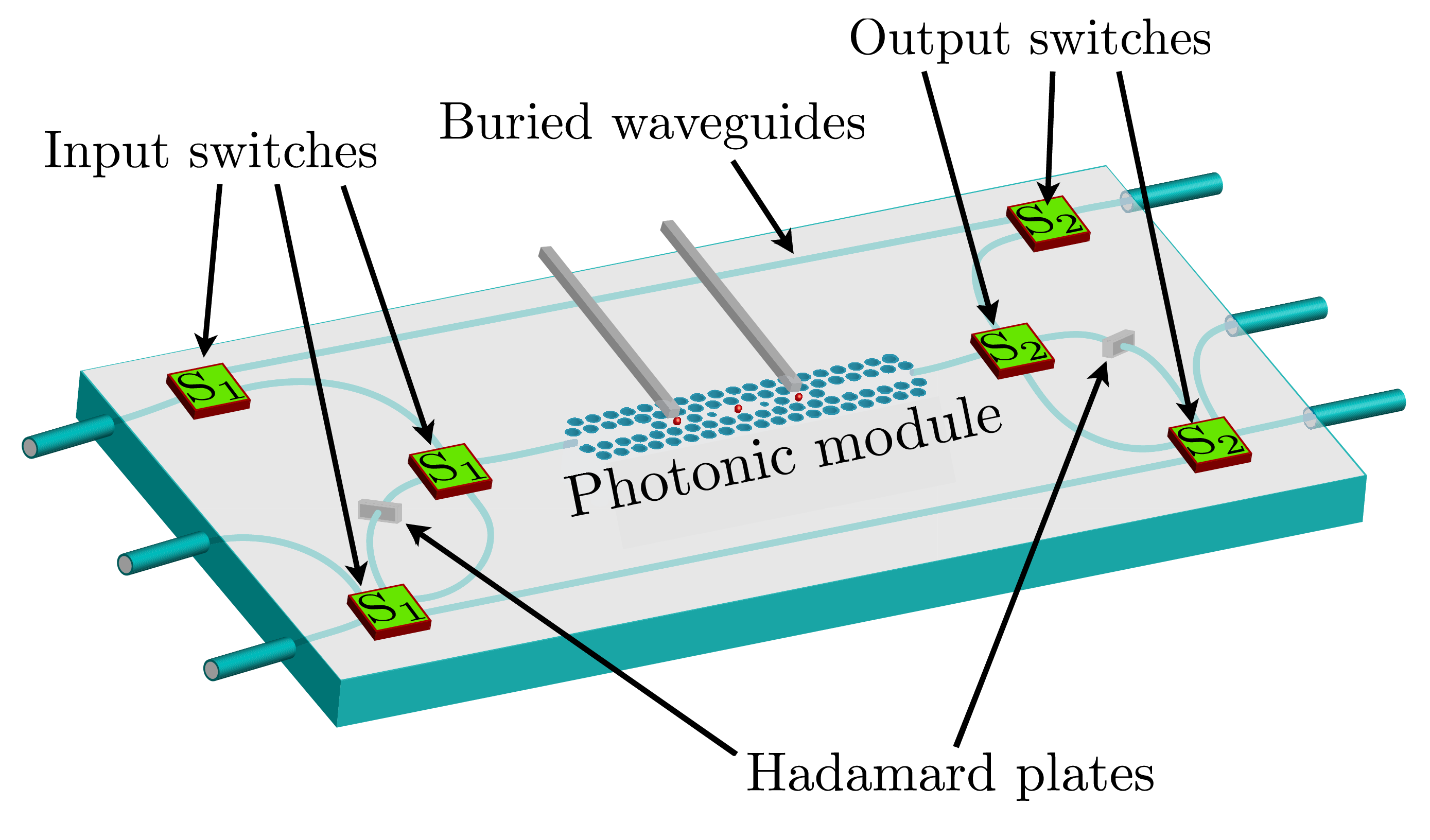}}
\end{center}
\vspace*{-10pt}
\caption{(Color online) A photonic module \cite{Devitt1} at the centre of a photonic chip, implemented here in a photonic bandgap structure. Photons are adiabatically loaded from the first Q-switch cavity into the module cavity which contains the atomic system with a differential coupling between photon polarization. Once the interaction is complete photons are out-coupled from the second Q-switch cavity into the right waveguide mode. The atomic qubit has laser control for initialization and measurement and each Q-switch has Stark shift controls.}
\label{figure:module1}
\end{figure}

\begin{figure}
\begin{center}
\resizebox{40mm}{!}{\includegraphics{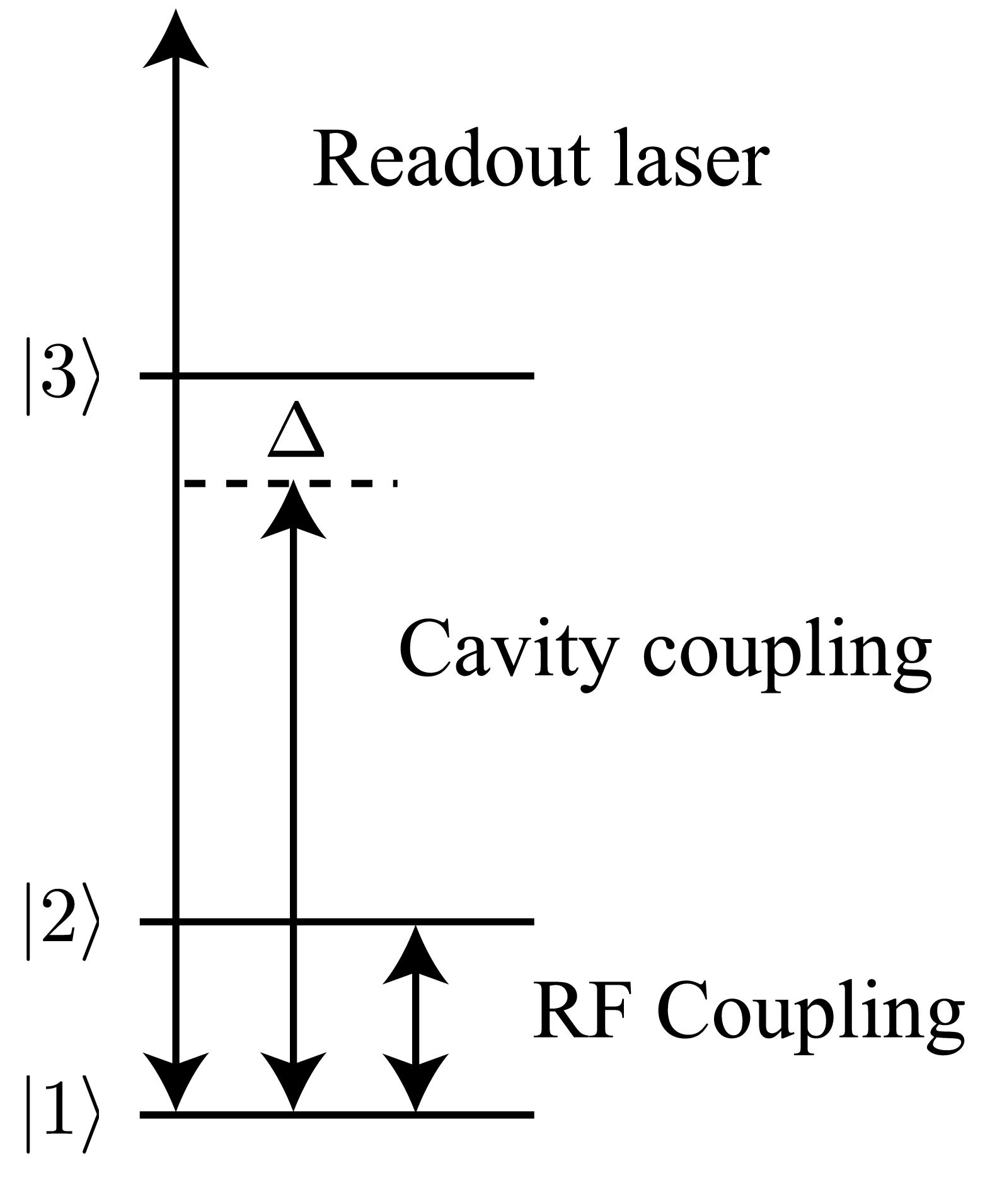}}
\end{center}
\vspace*{-10pt}
\caption{A three-level atomic system in the central cavity of the photonic module provides the non-demolition measurements at the heart of the scheme. The system is initialized in the state $\ket{1}$. With the resonant RF field, the system is pumped to the state $(\ket{1}+\ket{2})/\sqrt{2}$. The cavity mode is coupled to the $\ket{1}\rightarrow \ket{3}$ transition via the Hamiltonian $H=\beta a^{\dagger}a \sigma_z$, and we assume no dipole moment for the coupling from $\ket{2}$ to $\ket{3}$. Introducing a single photon into the cavity mode will induce a phase shift on the atomic state $\ket{1}$. The photon is controllably out-coupled from the cavity mode (via Q-Switched cavities) once the accumulated phase shift reaches $\pi$, hence the atomic system will oscillate between the $(\pm \ket{1} + \ket{2})/\sqrt{2}$ states with each sequential photon. After all photons that have passed through the system the RF pulse is applied again and the system measured in the $\{\ket{1},\ket{2}\}$ basis.}
\label{figure:cavity}
\end{figure}

Our aim is for a device that is effective, scalable, and relatively easy to fabricate. Photonic bandgap structures (PBS) and similar solid state strategies are therefore most appropriate. Figure \ref{figure:module1} illustrates the basic design of the PBS version of the photonic chip which contains the photonic module. Without loss of generality we can consider implementation using nitrogen vacancy (NV) centres in diamond \cite{Greentree2}. Three atomic qubits are incorporated into individual, coupled, high-Q cavities. The central atom-cavity system is the main component of the photonic module and the adjacent atom cavity systems act as one-atom Q-switches \cite{Greentree1} that allow adiabatic in- and out-coupling of a single photon into the central cavity.

The role of Q-switching to the efficient running of cavity-QED couplers is important, and warrants some extra discussion. To realise single atom, single photon coupling requires both a large photon intensity and long storage time, which implies high Q and small mode volume. If we just consider the issue of cavity Q then it is immediately obvious that without dynamic control, high Q implies small bandwidth, and hence fundamental limitations on the gate operation time. Worse still, there are potential problems with ringing - that is, oscillations in the photon intensity due to poorly matched pulses. One potential solution that breaks the time-bandwidth limitation is to dynamically vary the system properties~\cite{Xu1} - for example, to switch  from low Q while the photon pulse is entering, to high Q for atom-photon interaction, and then back to low Q again to outcouple the photon.

Although Q-switching is standard practice for classical lasers, there are few schemes that have been described that work with photonic bandgap structures and at the quantum level. Some examples of Q-switching schemes which appear to be feasible are the coupled cavity scheme of Ref.~\cite{Greentree1} which is our main focus here, the dynamic Stark shift scheme of Ref.~\cite{Fernee1}, and the EIT scheme of Ref.~\cite{Bermel1}. Estimates for the switching parameters from Ref.~\cite{Greentree1} suggest that the timescale for NV based photonics will be of order 100ns, and further optimizations suggest that shorter gate times of around 50ns will be possible in such systems~\cite{Su1}.
 
In addition to these control issues, the actual fabrication of devices such as that shown in Fig.~\ref{figure:module1} remains a major technological challenge. However, recent experimental results give cause for optimism: optical coherent manipulation of single color centers in diamond is now routine at room temperature \cite{Jelezko1}; strong coupling between a single color center and a microsphere cavity has been reported \cite{Park1}; nanofabrication techniques for diamond photonic crystals are being developed \cite{Wang1}; generation and transfer of photons on a photonic crystal chip has been demonstrated \cite{Englund2}; and already solid-state integrated optical approaches have proved useful for realizing non-deterministic quantum gates \cite{Politi1, Clark1}.

A single photon present in the central cavity mode must induce a non-destructive bit flip on the atomic qubit. This can be achieved in several ways \cite{Devitt1} but arguably the simplest is to exploit some of the previously demonstrated readout and control properties of nitrogen-vacancy color centers in diamond. Taking the usual approximation of the center as a three-level atom and ignoring the other transitions which are unimportant for our purposes, we assume that the ground states are coupled using a resonant RF field (derived from a field coil) at around 2.88 GHz (the zero magnetic field ground state splitting) which can be used to perform complete control of the ground state transitions \cite{Howard1}.  Spin selective readout of the ground states of individual nitrogen vacancy centers is by now routine via monitoring the induced fluorescence at 637nm from green laser excitation \cite{Jelezko1}. Coherent coupling of the optical transition has also been demonstrated \cite{Santori1}, although not yet at the one photon level, nor with cavities. Nonetheless, recent developments in diamond nano-fabrication \cite{Olivero1, Wang1} and design \cite{Hanic1} give considerable cause for optimism that such structures will soon be available.
 
The required atomic structure is therefore realized by a three-level atom in the central cavity in the $\Lambda$ configuration.  An RF field prepares the two ground states in a symmetric superposition state, and the cavity field couples one of the ground states to the excited state.  Note that in NV systems under the conditions for single-center readout \cite{Santori1}, only one of the ground states will have an allowed dipole transition to the excited state, which considerably eases the experimental burden. This is illustrated in Fig.~\ref{figure:cavity}. In the dispersive limit, this coupling will cause one of the atomic states to accumulate a phase shift depending on the total time the cavity mode is occupied.  The two Q-switches are used to control this interaction time  such that a $\pi$ phase shift is induced in the atomic system after which the photon is removed from the cavity and out-coupled to a waveguide. Utilizing this scheme we can effectively alter the Hamiltonian from $H = \beta a^{\dagger} a \sigma_z$ to $H' = \beta a^{\dagger} a \sigma_x$, where $\beta=-g^2 / \Delta$ is the usual off-resonance light shift for atom-cavity coupling $g$ and detuning $\Delta$, and so $\pi \Delta / g^2$ is the time required to induce the non-destructive atom-photon interaction.

In the PBS version of the module we assume that the coupling between atom and photon is polarization dependent - that is, only one component couples to the atomic system. For conceptual simplicity we assume that only the vertical component of polarization couples to the atomic system. In this case if a photon is prepared in the state $\ket{\pm} = (\ket{H}\pm\ket{V})/\sqrt{2}$ then the action of the module, $M$, is,
\begin{equation}
M\ket{+}\ket{\phi} = \ket{+}\ket{\phi}, \quad M\ket{-}\ket{\phi} = \ket{-}X\ket{\phi},
\end{equation}
for an arbitrary state $\ket{\phi} = \alpha\ket{0}+\ket{1}$ of the atomic system.  It is easy to check that this is completely equivalent to the transformation
\begin{equation}
M\ket{\psi}\ket{+} = \ket{\psi}\ket{+}, \quad M\ket{\psi}\ket{-} = X\ket{\psi}\ket{-},
\end{equation}
where the atomic qubit is prepared in the $\ket{\pm}$ state, the photon is in an arbitrary state $\ket{\psi} = \alpha\ket{H}+\beta\ket{V}$, and the bit flip now affects the state of the photon. This is the general action of the module. If we pass multiple photons through the system the transformation of a general $N$ photon state is,
\begin{equation}
\begin{array}{rcl}
M^{\otimes N} \ket{\Psi}_N\ket{0}&=&\frac{1}{\sqrt{2}} \ket{\Psi}_N\left[ \ket{+}+X^{\otimes N} \ket{-} \right] \\
\\
&=&\frac{1}{2}\left[ \ket{\Psi}_N+X^{\otimes N} \ket{\Psi}_N \right] \ket{0}\\
\\
&+&\frac{1}{2}\left[ \ket{\Psi}_N-X^{\otimes N} \ket{\Psi}_N \right] \ket{1}.
\end{array}
\end{equation}
That is, the action of the module is to project a train of $N$ photons into a $\pm 1$ eigenstate of the operator $X^{\otimes N}$.  The measurement outcome of the atomic system will determine the outcome of the projection, with local $Z$ operations used to switch between eigenstates.  This scheme for entangling photons is entirely deterministic and entangling many photons only requires sending them each individually through the module between initialization and measurement of the atomic qubit - there is no photon number dependance on the internal structure or operating dynamics of the module.

To prepare an $N$ photon stabilizer state \cite{Gottesman1}, such as a cluster state, each of the $N$ stabilizers that describe the state must be measured. As each of the stabilizers of an arbitrary $N$ photon state is an $N$-fold tensor product of the operators $\{I,X,Y,Z\}$, the ability to projectively measure the operator $X^{\otimes N'}$ for $N' \leq N$ and to apply local recovery operations is sufficient to stabilize an arbitrary state. This stabilizer measurement is precisely what the module allows. Furthermore, in most cases it is not necessary to apply local recovery operations immediately following each stabilizer measurement. Instead it is sufficient to store the result of the measurement in classical memory in a reference frame that is defined by some known Pauli rotation. Because of this, if more than one module is available photons can proceed to a stabilizer measurement at the second module before the outcome of the prior measurement is determined. 

\section{Cluster state preparation}
\label{CSP}

\subsection{Constant time preparataion}

\begin{figure*}
\begin{center}
\resizebox{30mm}{!}{\includegraphics{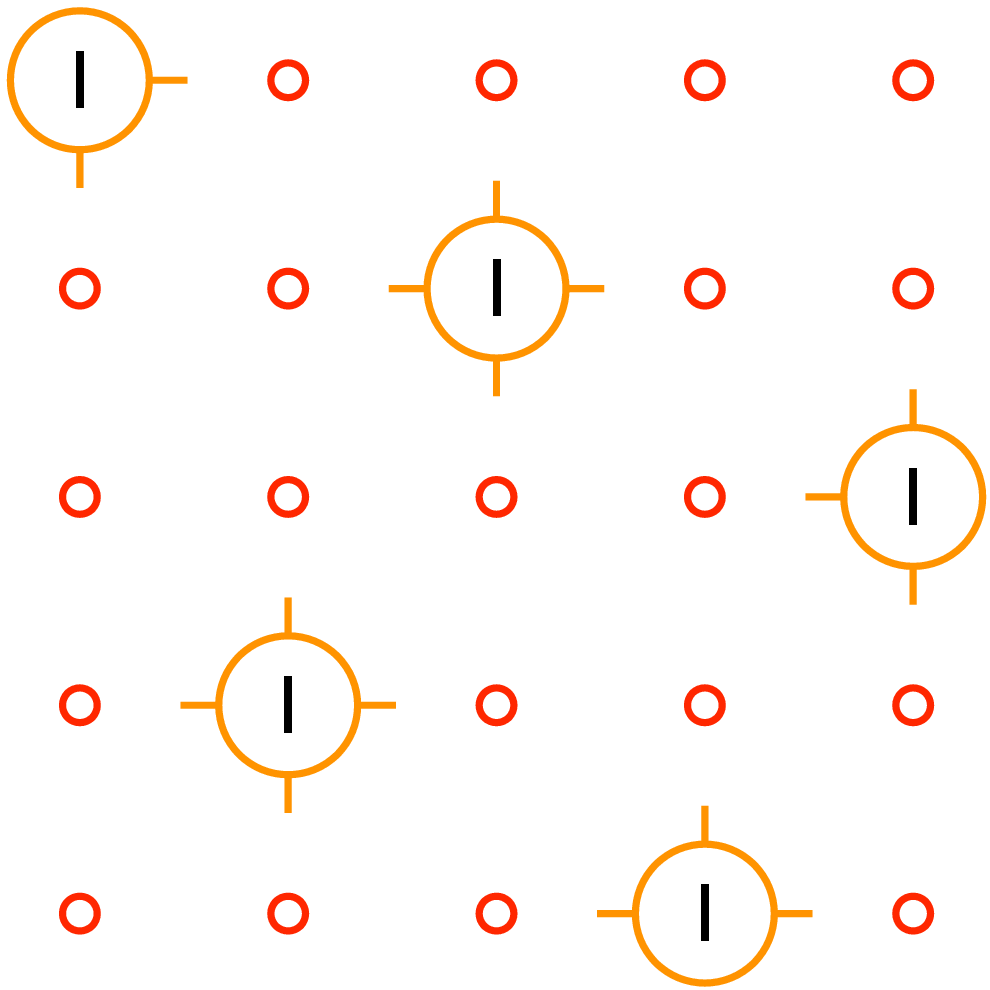}}
\hspace*{15pt}
\resizebox{30mm}{!}{\includegraphics{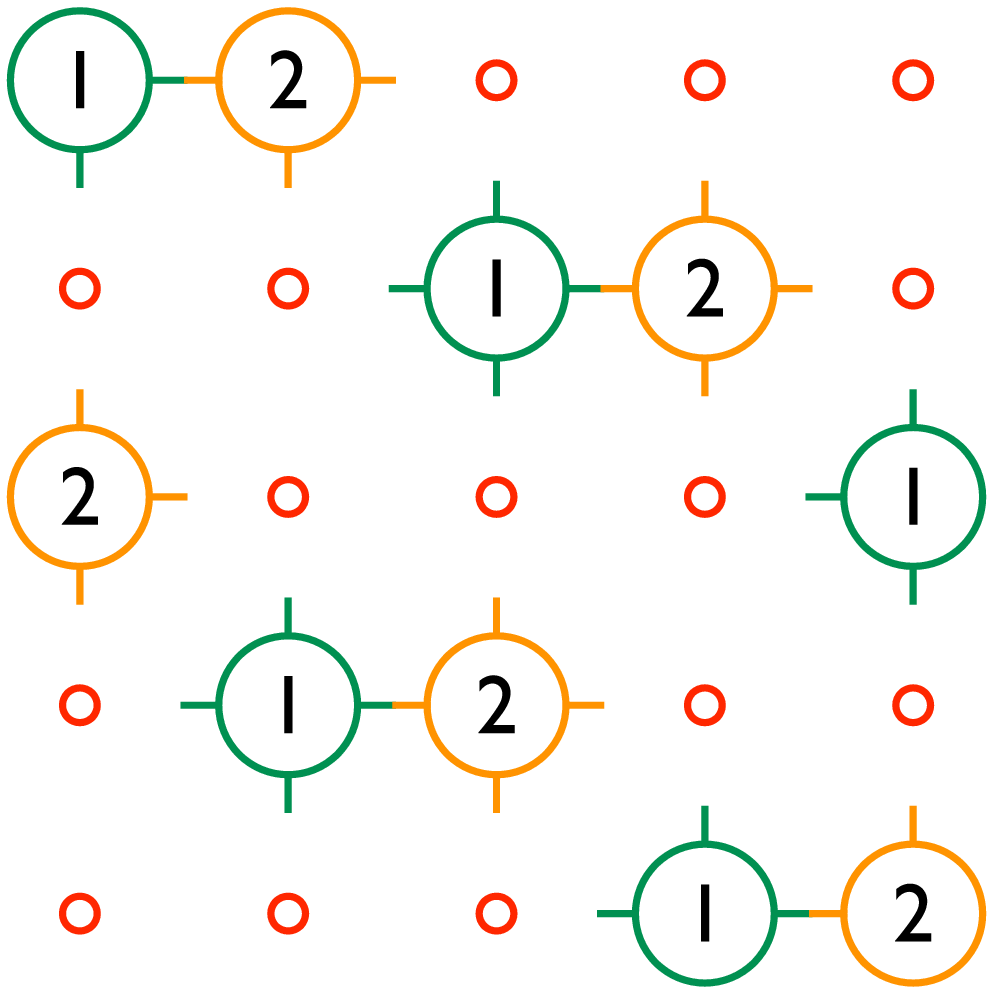}}
\hspace*{15pt}
\resizebox{30mm}{!}{\includegraphics{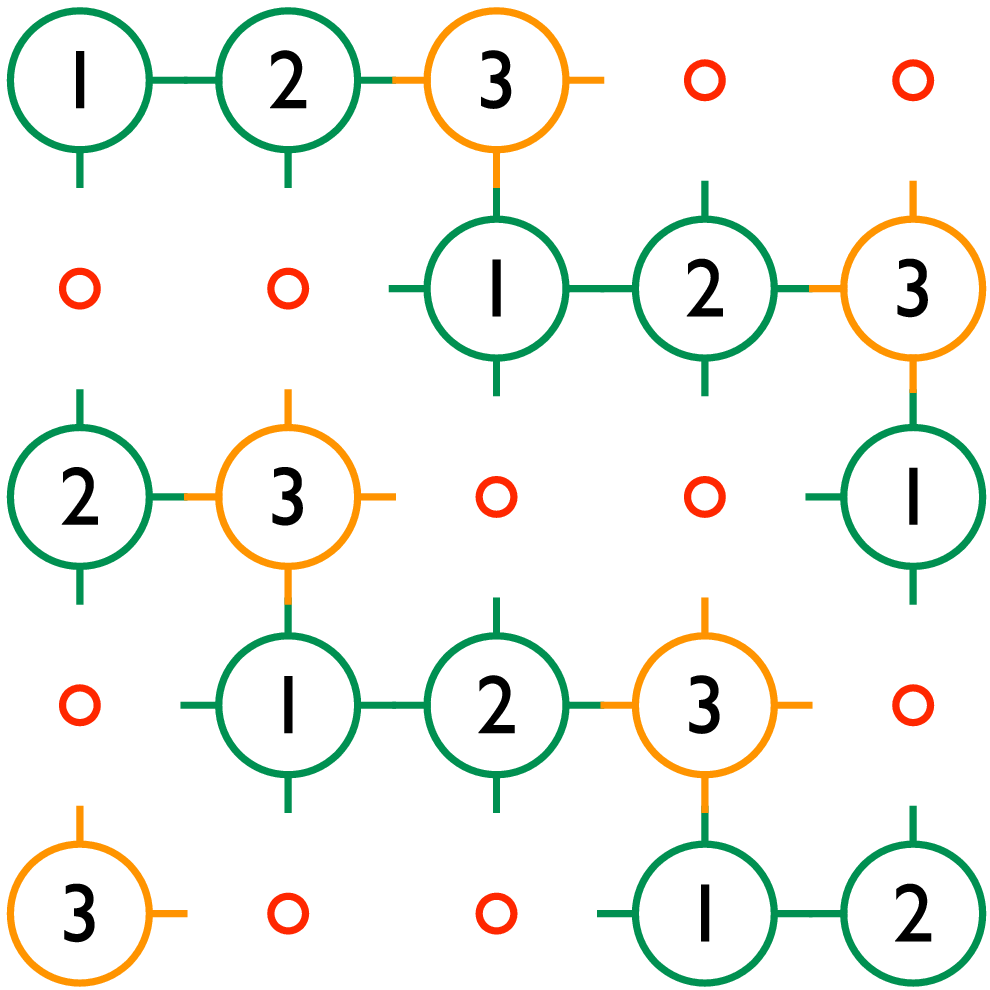}}
\hspace*{15pt}
\resizebox{30mm}{!}{\includegraphics{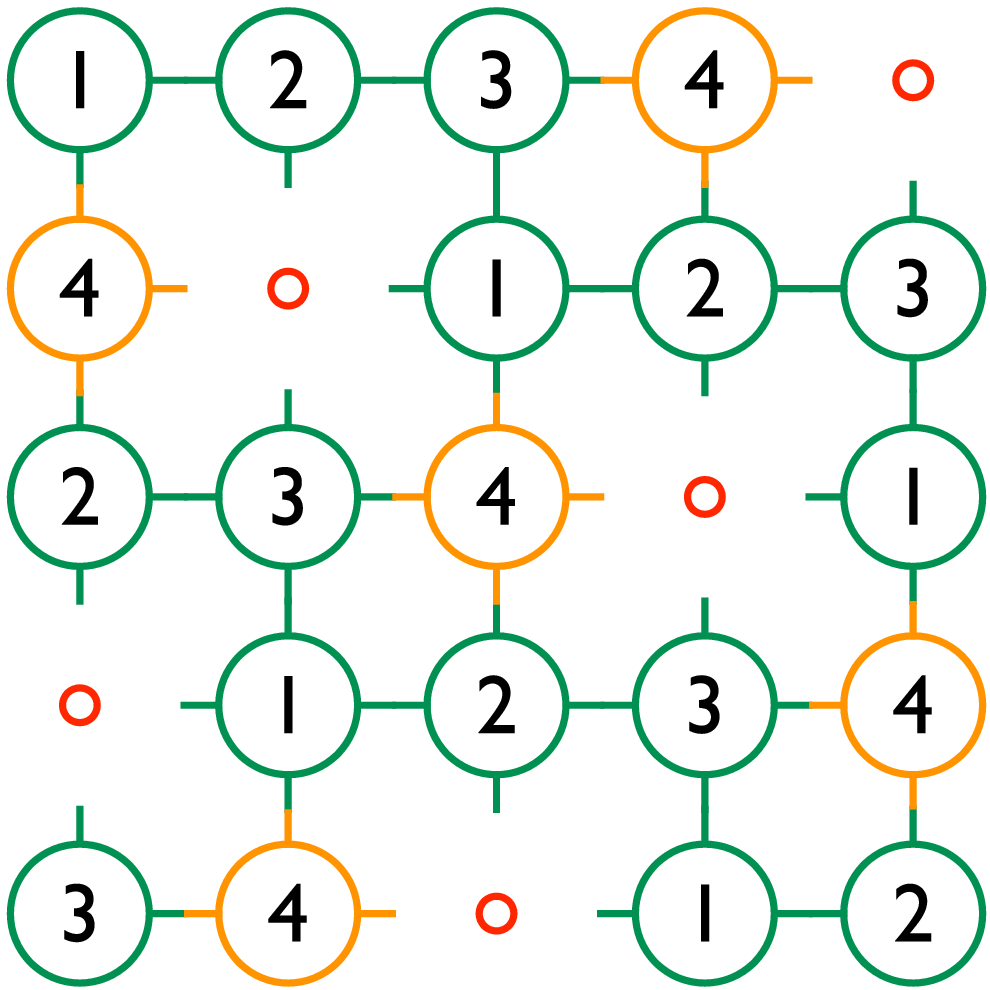}}
\hspace*{15pt}
\resizebox{30mm}{!}{\includegraphics{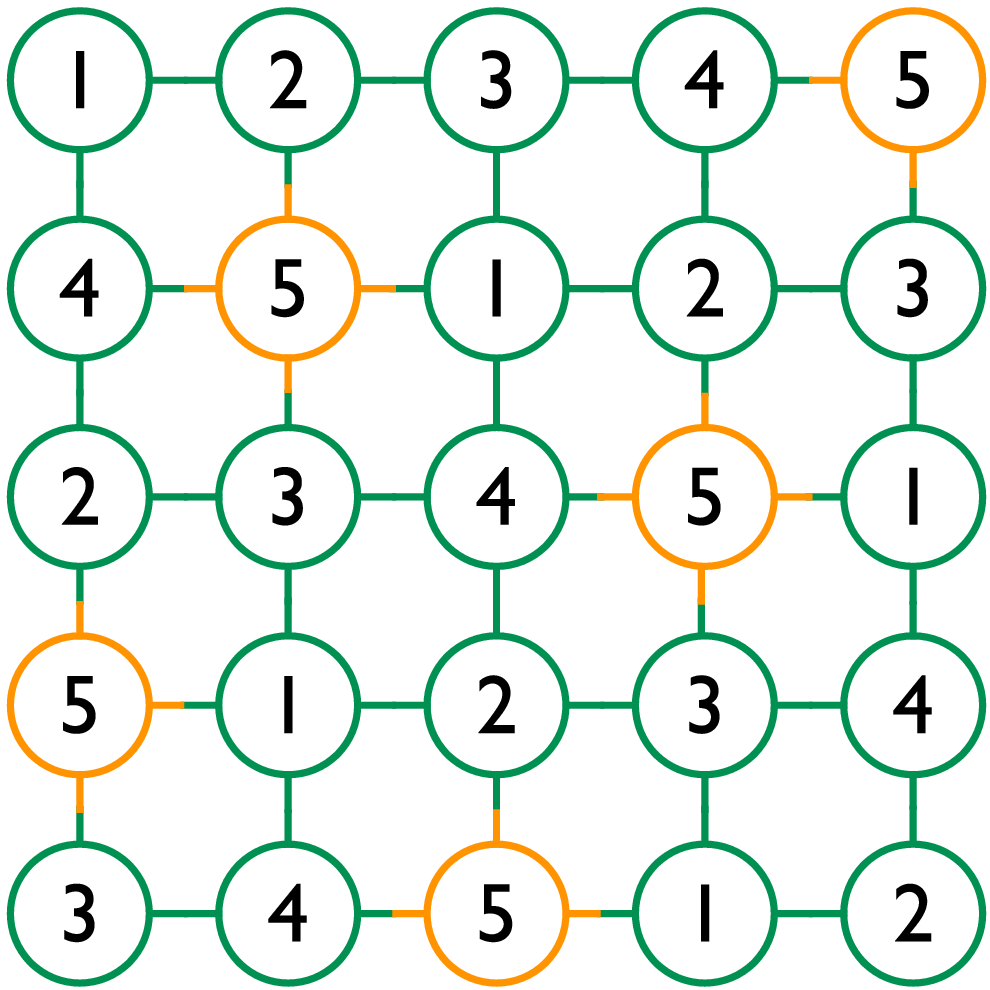}}
\end{center}
\vspace*{-10pt}
\caption{(Color online) Preparation of a two-dimensional cluster state in five time steps. Each vertex of the lattice represents a photon in the cluster and vertices are numbered to indicate the time step during which its associated stabilizer operator (see Eq.~\ref{stabilizer}) is measured. Green vertices represent stabilizer operators that have been measured, orange vertices represent stabilizer operators that are being measured, and red vertices represent stabilizer operators that are yet to be measured. By extending this pattern vertically and horizontally a larger cluster can be prepared.}
\label{figure:prep}
\end{figure*}

\begin{figure*}[t!]
\begin{center}
\resizebox{34mm}{!}{\includegraphics{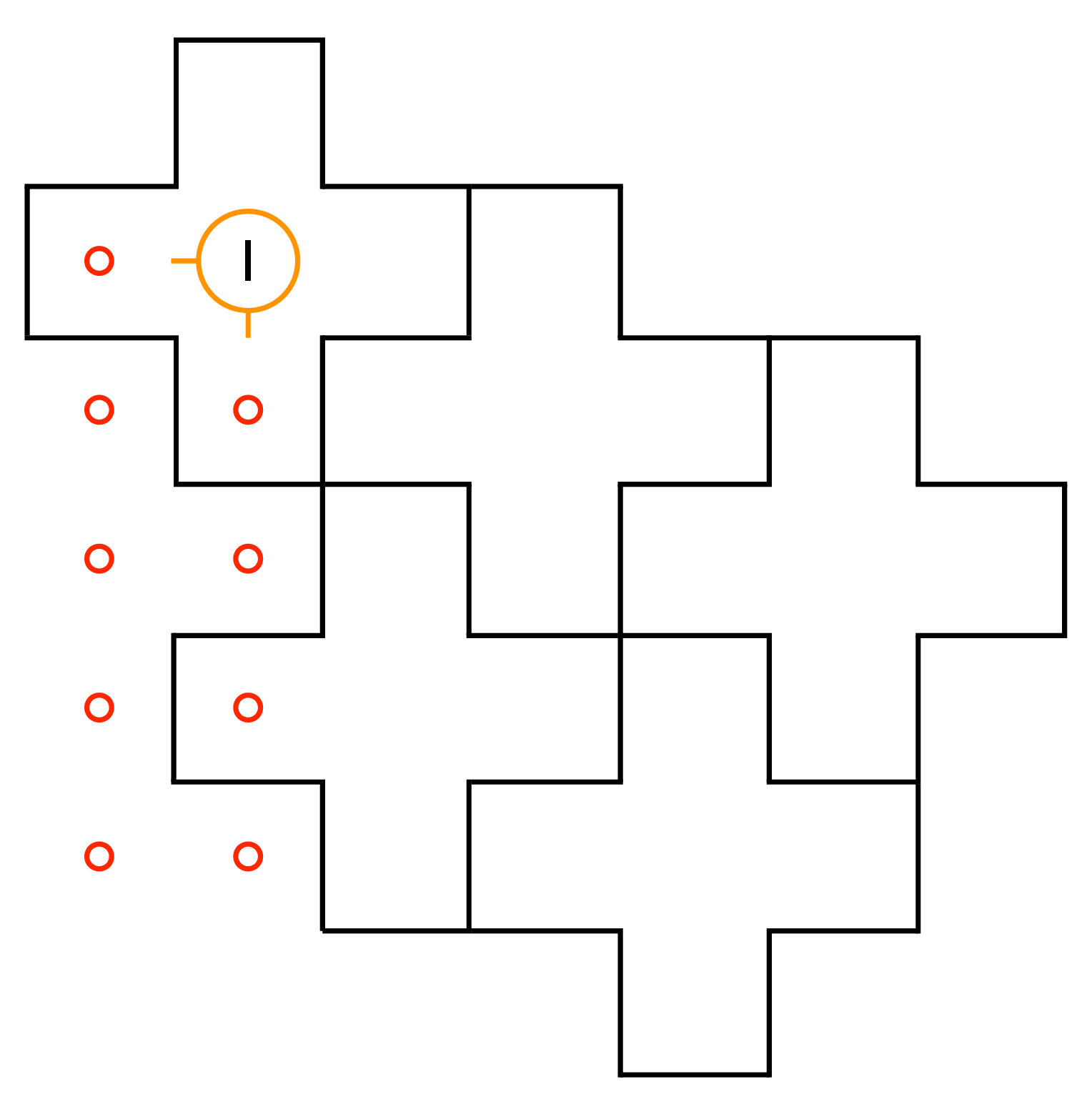}}
\hspace*{1pt}
\resizebox{34mm}{!}{\includegraphics{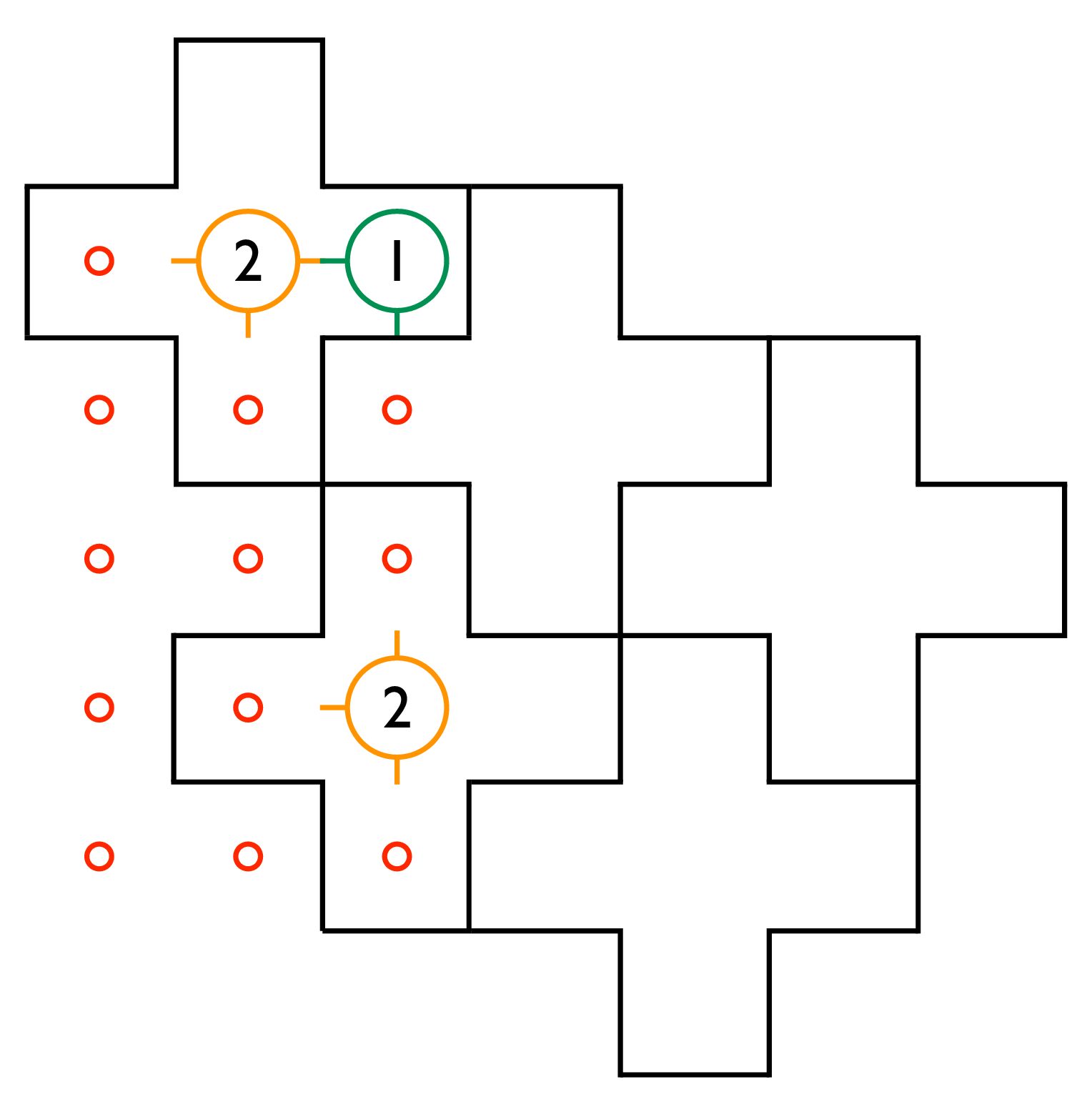}}
\hspace*{1pt}
\resizebox{34mm}{!}{\includegraphics{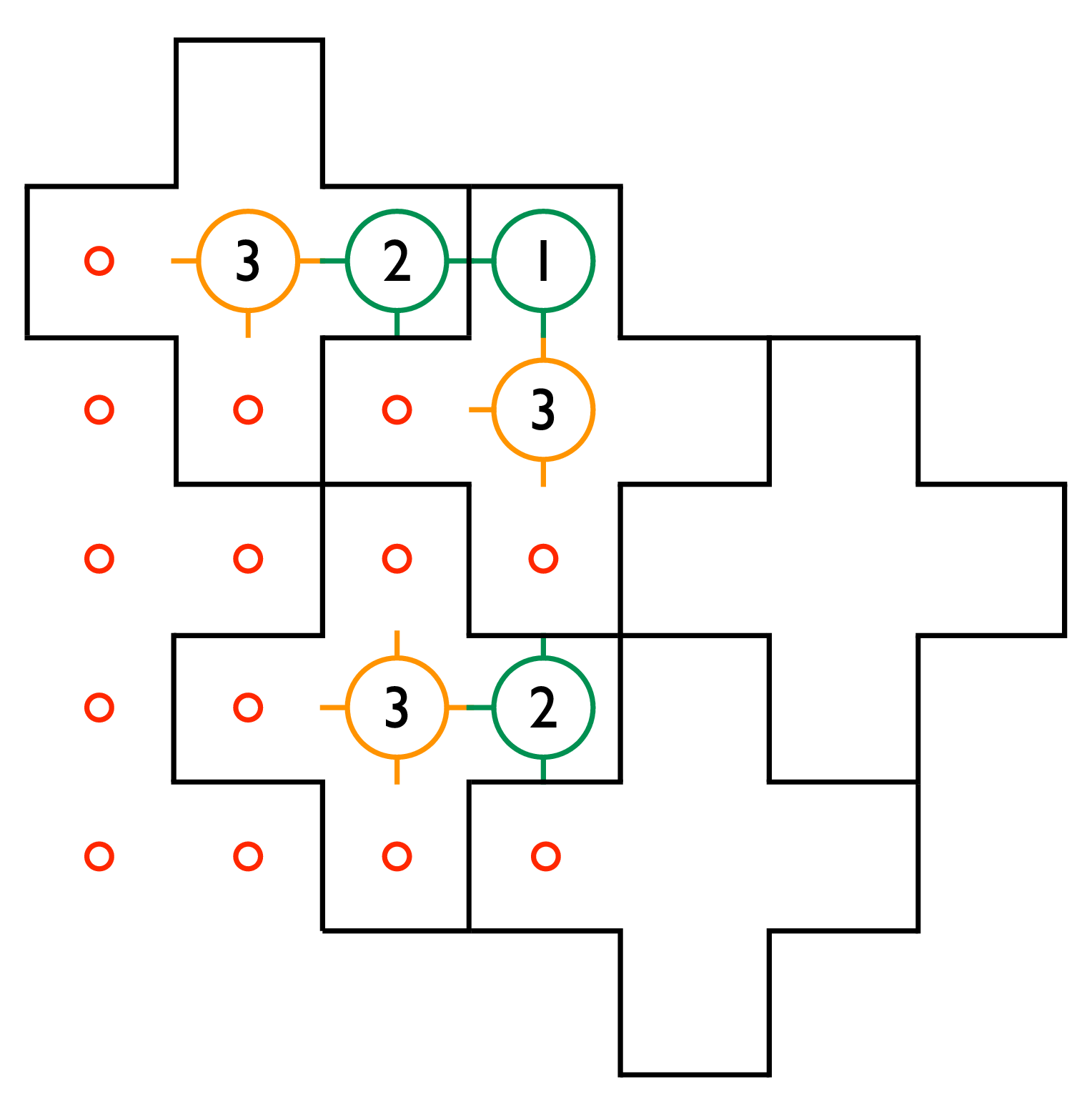}}
\hspace*{1pt}
\resizebox{34mm}{!}{\includegraphics{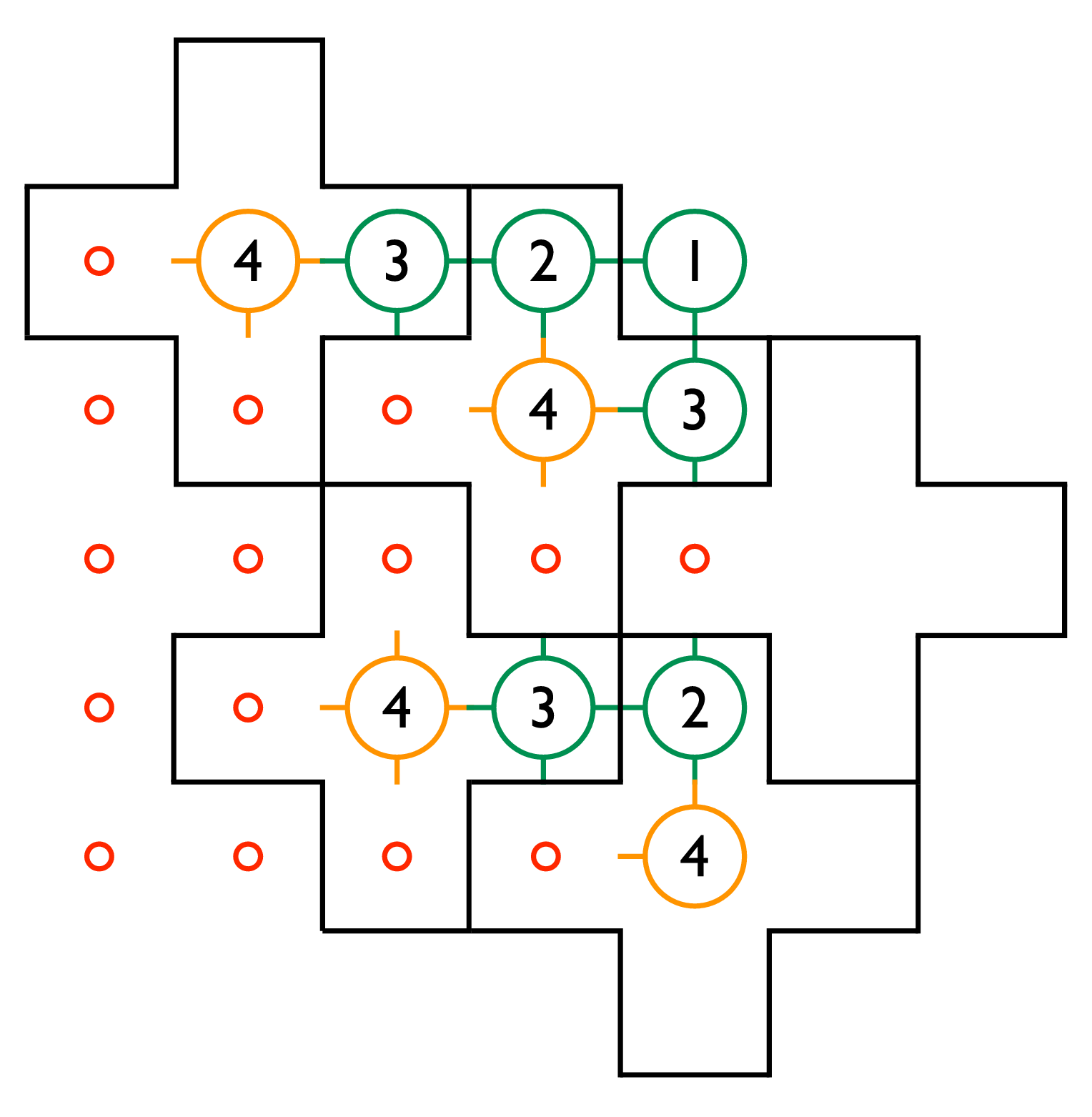}}
\hspace*{1pt}
\resizebox{34mm}{!}{\includegraphics{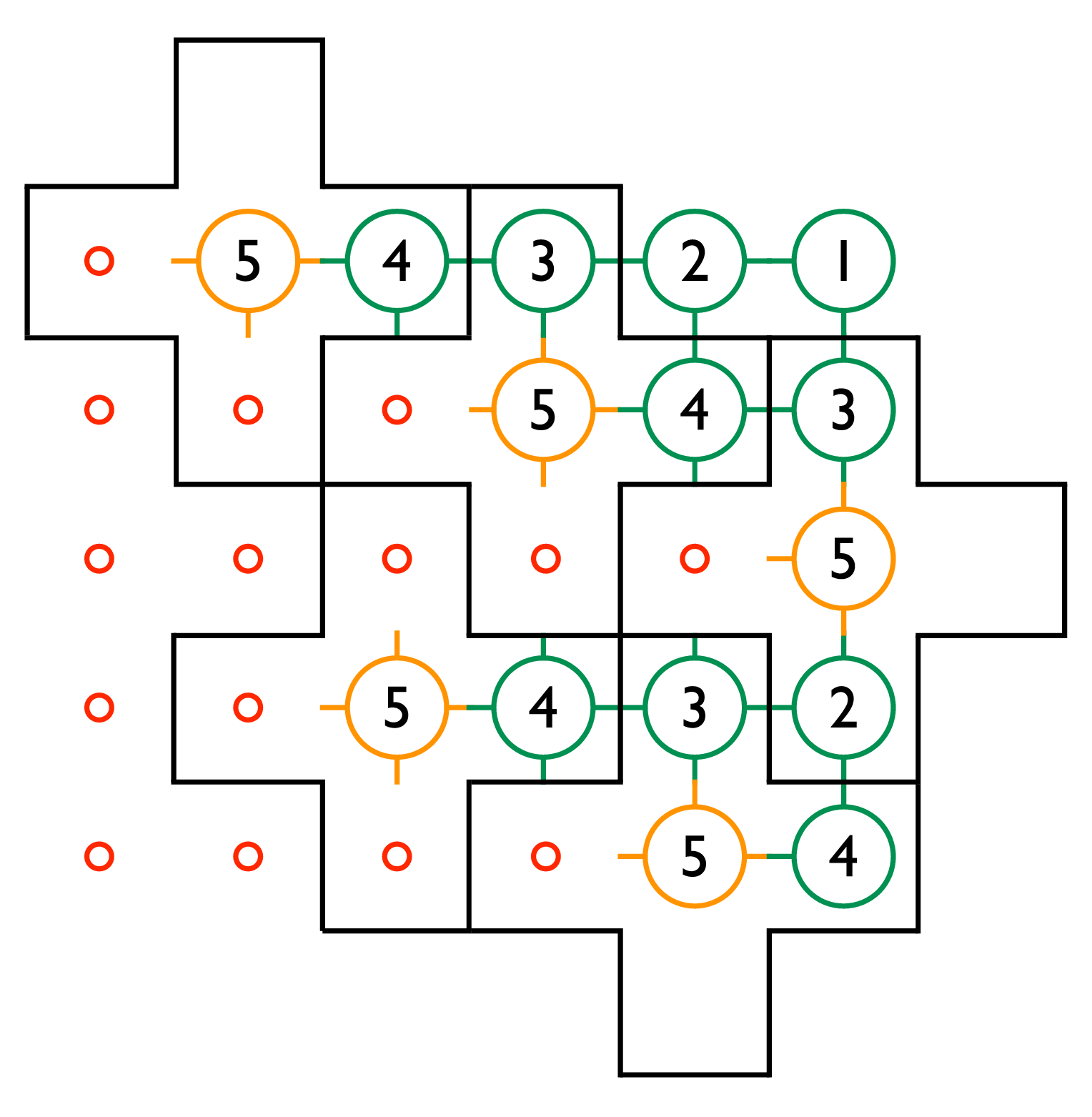}}
\end{center}
\vspace*{-10pt}
\caption{(Color online) Continuous preparation of a two-dimensional cluster state by the synchronous network. Each cross represents a photonic chip that will measure a stabilizer of the form of Eq.~\ref{stabilizer} whenever a photon occupies the central vertex. Other symbolic conventions are as in Fig.~\ref{figure:prep}. In every time step after the first four a new column of the two-dimensional cluster state is prepared. Unentangled photons can enter from the left indefinitely and the network can be extended vertically to prepare a larger cluster.}
\label{figure:prep2}
\end{figure*}

An $N$ photon two-dimensional cluster state is stabilized by $N$ stabilizers of the form
\begin{equation}
Z^{i-1,j}Z^{i,j+1}X^{i,j}Z^{i+1,j}Z^{i,j-1},
\label{stabilizer}
\end{equation} 
where $i,j$ are the co-ordinates of the $N$ photons that are topologically arranged on a two-dimensional square lattice. For any photon $i,j$ that does not have four nearest neighbors (any photon that is on an edge or a corner of the lattice) the associated stabilizer retains the form of Eq.~\ref{stabilizer} but excludes the operator(s) associated with the missing neighbor(s). Therefore, a cluster state can be prepared by performing $N$ stabilizer measurements that involve at most five photons each.

With only one module available, the cluster state could be prepared in $N$ time steps by sequentially measuring each stabilizer.  This is undesirable as photonic routing and storage would be non-trivial. With more modules available it would be possible to perform stabilizer measurements in parallel (using one module per measurement), however, as a single photon cannot be involved in more than one measurement simultaneously, the required measurements must be carefully arranged to achieve maximum parallelism. Figure \ref{figure:prep} illustrates this arrangement for the preparation of a 25 photon cluster. To prepare a larger cluster state, the pattern in Fig.~\ref{figure:prep} can be extended to cover the entire cluster, thus enabling an $N$ photon two-dimensional cluster state to be prepared in five time steps if $N/5$ modules are available.

The ability to prepare a cluster state in constant time may be beneficial for small computational tasks but for a large instance of an error corrected algorithm, for which a cluster of millions of qubits may be required, it is not ideal. Because the cluster will be consumed slowly during the execution of the algorithm by measurement, many photons would need to be stored for a significant length of time. Photonic storage not only presents a significant engineering challenge but will also increase the likelihood that the photons in the cluster will decohere or be lost before they are measured. To minimize photon storage it is necessary to continuously produce the cluster at a rate that is matched to the rate of consumption.

\subsection{Continuous synchronous preparation}

The pattern of stabilizer measurements in Fig.~\ref{figure:prep} can be realized by passing lines of photons through a network of static photonic chips, where each photonic chip comprises a photonic module and a simple optical network. This is illustrated schematically in Fig.~\ref{figure:prep2} for a cluster that has five rows: Unentangled photons enter the network from the left and travel horizontally. Any time when a photon is in the center of a chip a measurement of a stabilizer of the form of Eq.~\ref{stabilizer} is performed involving all photons in that chip. After the first four time steps a new vertical column of the cluster is prepared in every subsequent step. Unentangled photons can enter from the left indefinitely and the network can be extended vertically so that, provided one module is available for every row of the cluster, the rate of preparation of columns in the cluster is constant and equal for a cluster of any size.

\begin{figure}
\begin{center}
\resizebox{65mm}{!}{\includegraphics{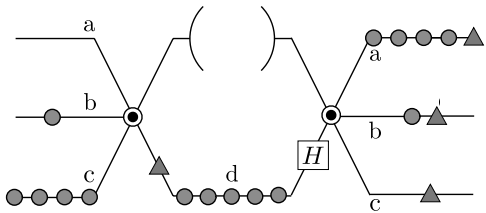}}
\end{center}
\vspace*{-10pt}
\caption{Photonic chip for the synchronous network. Each shaded circle represents a buffer that effects a delay of $\delta_t$ and each shaded triangle represents a buffer that effects a delay of $\delta_t^{\prime}$. The input and output switches are operated according to Tab.~\ref{table:routing1}. Note that line d is a feedback loop that can return photons to the input switch. Before they exit the module photons that entered along lines a and c are delayed by $5\delta_t+\delta_t^{\prime}$ and photons that entered along line b are delayed by $15\delta_t+3\delta_t^{\prime}$.}
\label{figure:chip1}
\end{figure}

For the network to function as illustrated in Fig.~\ref{figure:prep2} each photonic chip is required to switch in and out up to three photons simultaneously. If we choose as our unit of time the time required for an atom cavity interaction in the photonic module, $\delta_t$, and if we assume that additional time, $\delta_t^{\prime}$, is required for measurement of the atomic qubit, then every $5\delta_t+\delta_t^{\prime}$ a new photon is sent along each row. The photons entering each module are first staggered by buffers as illustrated in Fig.~\ref{figure:chip1}. Photons are then routed through the photonic module by switching as described in Tab.~\ref{table:routing1}. Note that incoming photons are supplemented by up to two photons that are held over in the delay line (by buffers) from previous time steps - incoming photons plus those already in the delay line are all involved in a stabilizer measurement. Tab.~\ref{table:routing1} indicates which photons are held over in the delay line for subsequent stabilizer measurement(s). The rate of preparation of the cluster state is limited by the rate at which photons enter the network: one column is prepared every $5\delta_t+\delta_t^{\prime}$ for a cluster of any size.

\begin{table}
\begin{center}
\begin{tabular}{|c|c|}
$I$ & $O$ \\
\cline{1-2}
$a \rightarrow cav$ & $-$ \\
$b \rightarrow cav$ & $cav \rightarrow a$ \\
$d \rightarrow cav$ & $cav \rightarrow d$ \\
$d \rightarrow cav$ & $cav \rightarrow d$ \\
$c \rightarrow cav$ & $cav \rightarrow b$ \\
$-$ & $cav \rightarrow c$
\end{tabular}
\vspace*{4pt}
\caption{Switching table for the synchronous network. The left-hand and right-hand columns give the switching settings for the input and output switches respectively of the photonic chip in Fig.~\ref{figure:chip1}. The labels $a$, $b$, $c$, $d$ refer to the lines in Fig.~\ref{figure:chip1} and $cav$ refers to the cavity in the photonic module. The order of the switching is from top to bottom, where the setting in each row is held for duration $\delta_t$, except for the final setting which is held for duration $\delta_t^{\prime}$ during which the atomic qubit is measured and after which the switching pattern repeats.}
\label{table:routing1}
\end{center}
\end{table}

Though the synchronous network is conceptually simple - each chip is identical, all chips are operated synchronously, and only one photonic chip is required for each row of the cluster state - the main disadvantage of this network is the requirement for buffering in the photonic chip. Buffering is expected to be a source of decoherence and will increase the difficulty of fabrication and characterization of each photonic chip.

\subsection{Continuous asynchronous preparation}

\begin{figure}
\begin{center}
\vspace*{10pt}
\resizebox{50mm}{!}{\includegraphics{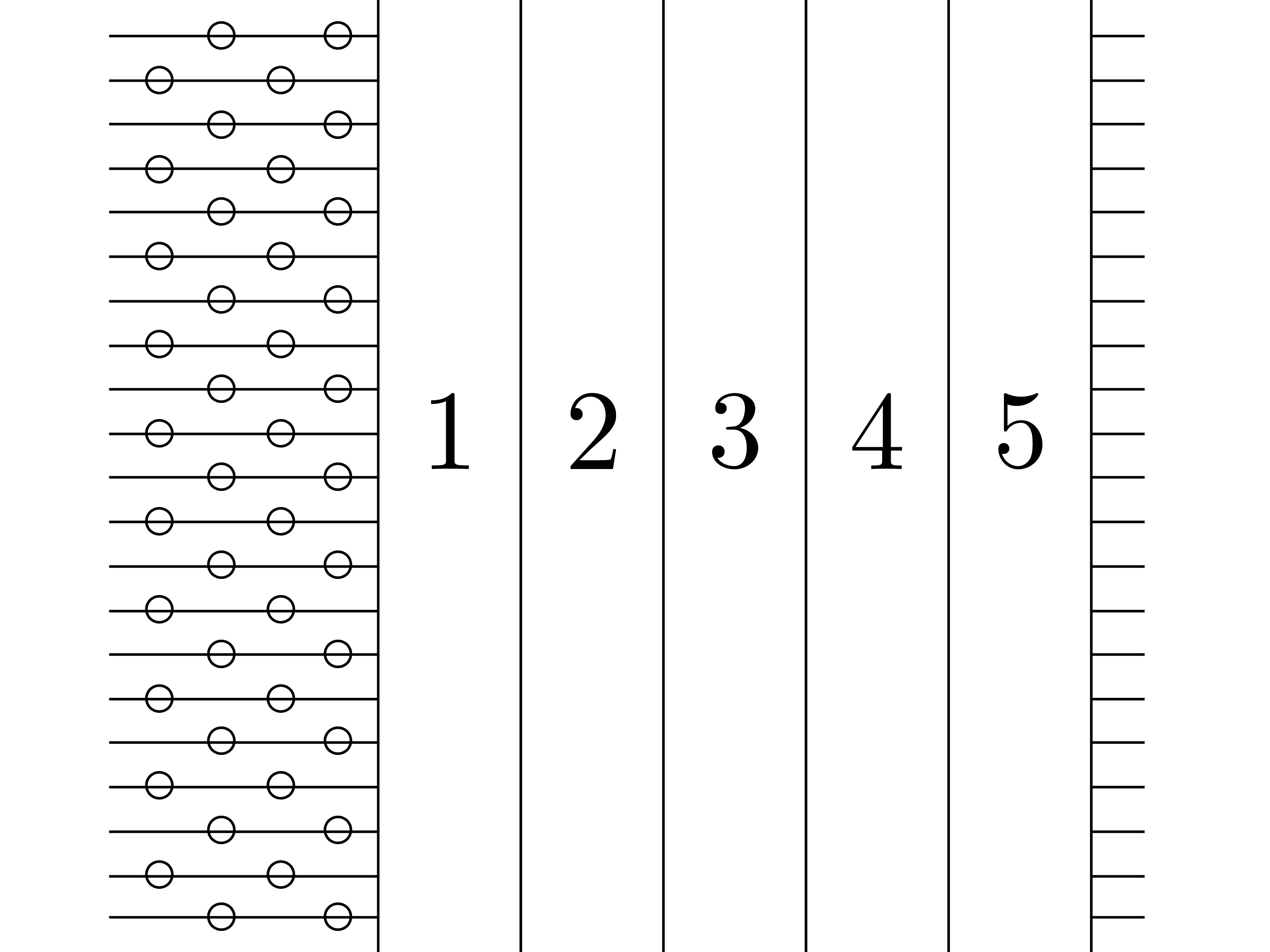}}
\end{center}
\vspace*{-10pt}
\caption{Asynchronous network for the continuous preparation of a two-dimensional cluster state. Every column is an identical set of photonic chips as illustrated in Fig.~\ref{figure:column}. The time interval between input photons on each row is $2\delta_t$ and adjacent rows are staggered by $\delta_t$.}
\label{figure:columns}
\end{figure}

\begin{figure}
\begin{center}
\vspace*{10pt}
\resizebox{45mm}{!}{\includegraphics{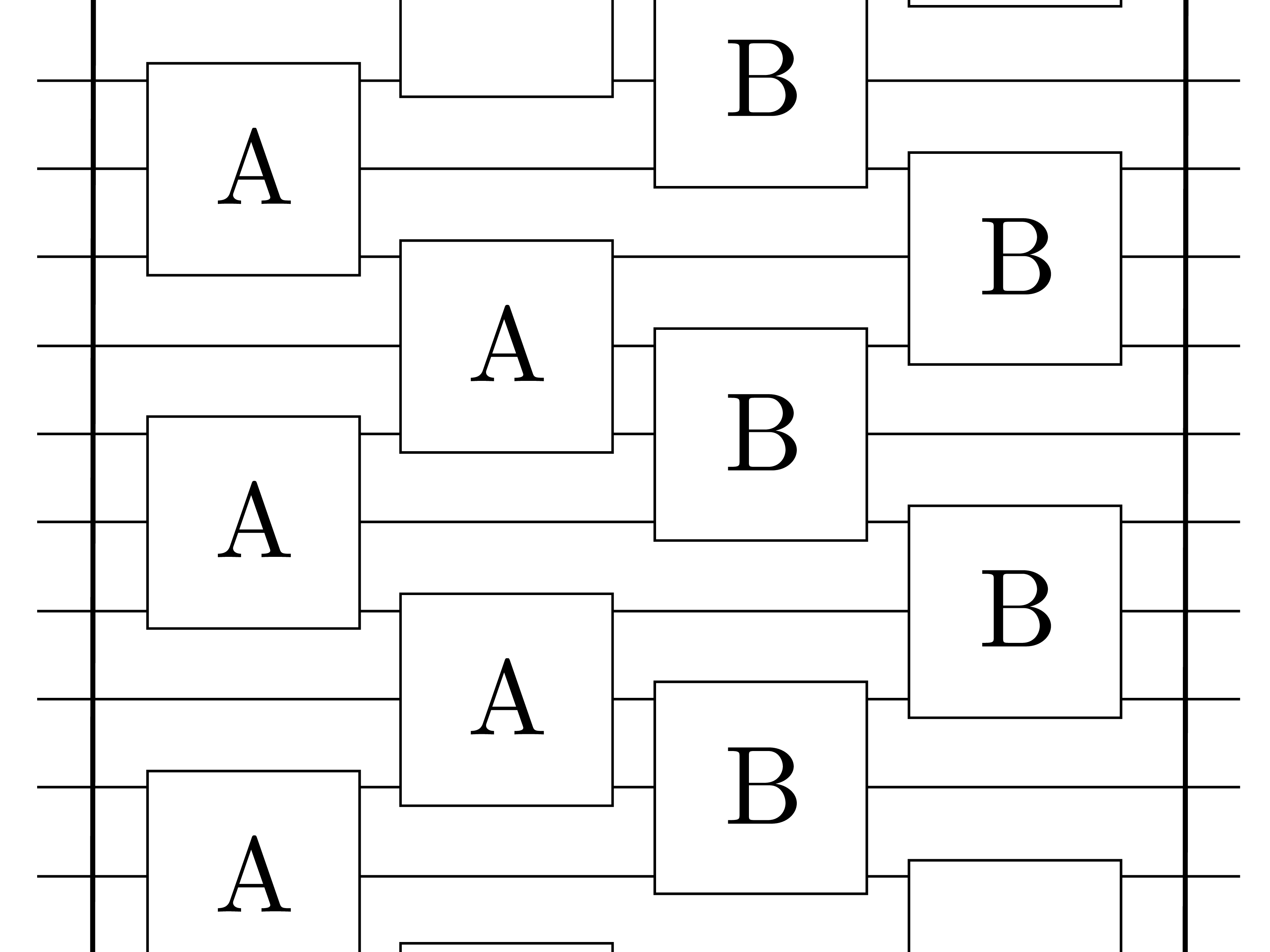}}
\end{center}
\vspace*{-10pt}
\caption{One of five identical columns in the asynchronous network of Fig.~6. Each box represents a photonic chip as shown in Fig.~9, all of which are the same. The timing of their operation can be found in Tab.~II.}
\label{figure:column}
\end{figure}

To eliminate the need for buffering it is necessary to increase the number of photonic chips. In the asynchronous network the photonic chips are arranged in five columns with one photonic chip centered on every row in each of the five columns. The precise arrangement of photonic chips is illustrated in Figs.~\ref{figure:columns} and \ref{figure:column}. Note that every photonic chip in every column is physically identical and that the layout of the chips in each of the five columns is identical. As before, unentangled photons enter the network from the left and travel horizontally but, unlike in the synchronous network, in the asynchronous network photons are not delayed anywhere except for when they interact in the cavity of a photonic module for time $\delta_t$. Also note that input photons are now staggered as is illustrated in Fig.~\ref{figure:columns}, where the time between photons on any row is now $2\delta_t$ and adjacent rows are staggered by $\delta_t$. 

To understand how the asynchronous network functions it is instructive to consider the passage of a single photon through the network. As a photon passes though the five columns it will be routed to the cavity of a photonic module only once per column - it bypasses every other photonic module in the column. Each time the photon is routed to a cavity it is involved in a single stabilizer measurement of the form of Eq.~\ref{stabilizer}. For each of the five stabilizer measurements (one per column) the photon is in a different position in the stabilizer - that is, once the photon exits the network it will have been the top, bottom, left, right, and center photon in five different stabilizers as required. As the photon is only delayed by its interaction with the cavity the total time spent inside the entire network is $5\delta_t$.

\begin{figure}
\begin{center}
\resizebox{80mm}{!}{\includegraphics{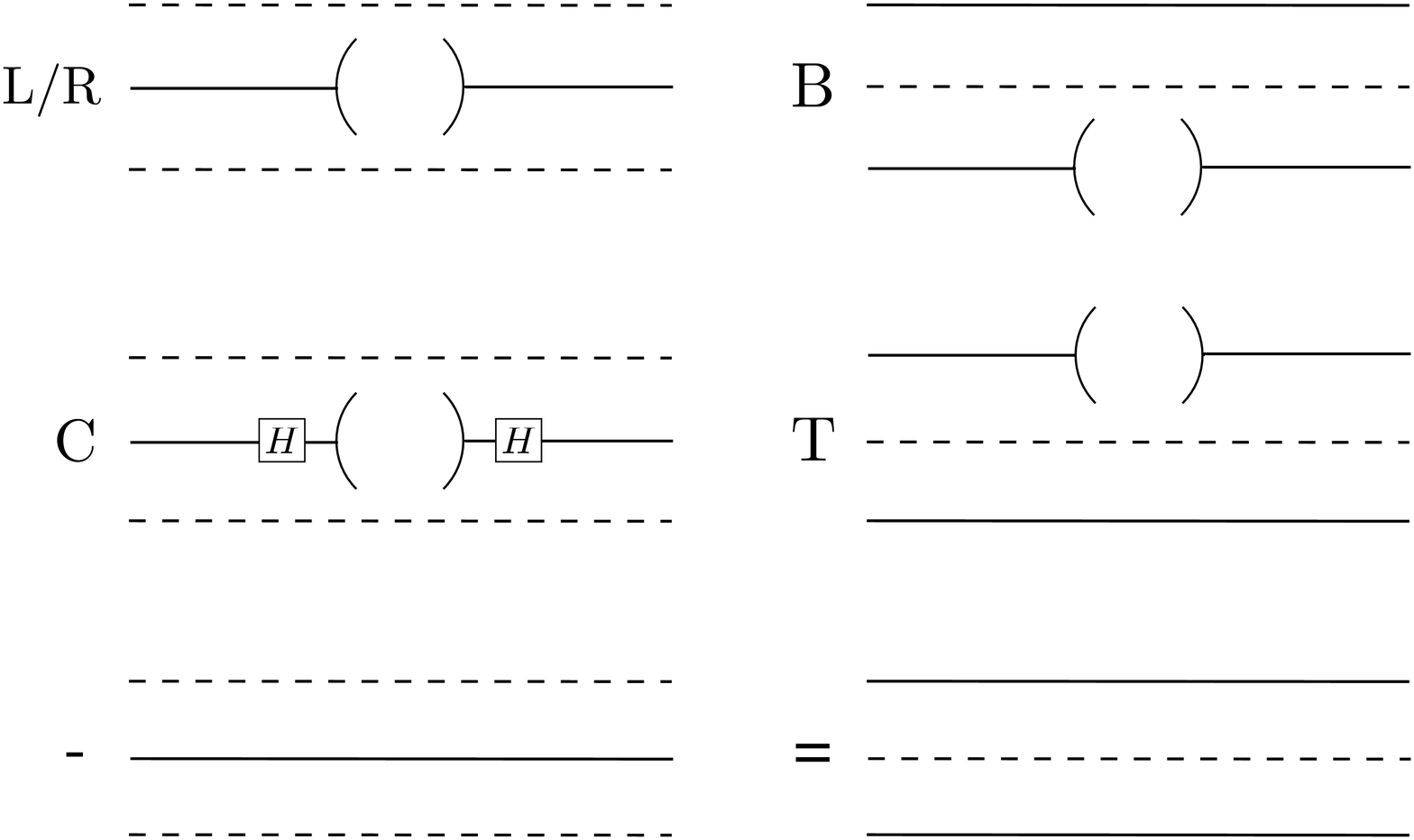}}
\end{center}
\vspace*{-15pt}
\caption{The six switching settings required for the photonic chip in the asynchronous network. Solid and broken lines indicate the presence and absence of a photon respectively. Parentheses indicate that the photon should be switched to the cavity of the photonic module, no parentheses indicate that the photon should be switched to the bypass line, and $H$ indicates that a Hadamard waveplate is required. These six settings can be achieved using the photonic chip in Fig.~\ref{figure:chip2}.}
\label{figure:chip2req}
\end{figure}

\begin{table}
\begin{center}
\begin{tabular}{|c|c|c|c|c|c|c|c|c|c|c|}
$$ & $1$ & $2$ & $3$ & $4$ & $5$ & $6$ & $7$ & $8$ & $9$ & $10$ \\
\cline{1-11}
$A_1$ & $R$ & $T$ & $C$ & $B$ & $L$ & $=$ & $-$ & $=$ & $-$ & $=$ \\
$B_1$ & $=$ & $-$ & $=$ & $-$ & $=$ & $R$ & $T$ & $C$ & $B$ & $L$ \\
$A_2$ & $=$ & $-$ & $R$ & $T$ & $C$ & $B$ & $L$ & $=$ & $-$ & $=$ \\
$B_2$ & $B$ & $L$ & $=$ & $-$ & $=$ & $-$ & $=$ & $R$ & $T$ & $C$ \\
$A_3$ & $-$ & $=$ & $-$ & $=$ & $R$ & $T$ & $C$ & $B$ & $L$ & $=$ \\
$B_3$ & $T$ & $C$ & $B$ & $L$ & $=$ & $-$ & $=$ & $-$ & $=$ & $R$ \\
$A_4$ & $L$ & $=$ & $-$ & $=$ & $-$ & $=$ & $R$ & $T$ & $C$ & $B$ \\
$B_4$ & $=$ & $R$ & $T$ & $C$ & $B$ & $L$ & $=$ & $-$ & $=$ & $-$ \\
$A_5$ & $C$ & $B$ & $L$ & $=$ & $-$ & $=$ & $-$ & $=$ & $R$ & $T$ \\
$B_5$ & $=$ & $-$ & $=$ & $R$ & $T$ & $C$ & $B$ & $L$ & $=$ & $-$ \\
\end{tabular}
\vspace*{4pt}
\caption{Routing table for the asynchronous network. The ten rows give the switching settings for the ten chip types in the network, where the subscript denotes the network column number and A and B the chip type within that column. The six switching settings can be correlated to the settings in Fig.~\ref{figure:chip2req} where the letters are chosen to indicate that all photons that are switched to the cavities of the photonic modules in any column in any time step are either the right, top, center, bottom, or left photons in stabilizers of the form of Eq.~\ref{stabilizer}. The order of the switching is from left to right, where each row of the table is a unique time step and time steps are separated by $\delta_t$. Switching to the photonic module of A and B chips is alternated in each column to allocate time for readout of the atomic qubit in the photonic module (which occurs when switched to $-$ or $=$). Note that the routing in each column is identical except for an offset of $2\delta_t$ and that the routing table repeats after every $10\delta_t$.}
\label{table:routing2}
\end{center}
\end{table}

\begin{figure}
\begin{center}
\resizebox{65mm}{!}{\includegraphics{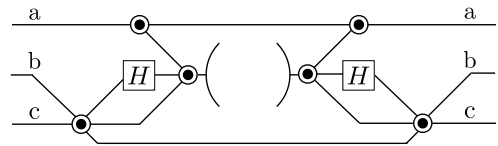}}
\end{center}
\vspace*{-15pt}
\caption{Photonic chip for the asynchronous network, also illustrated in Fig.~\ref{figure:module1}. See Tab.~\ref{table:routing2} for details of the routing of photons through the photonic module.}
\label{figure:chip2}
\end{figure}

\begin{figure*}
\begin{center}
\resizebox{34mm}{!}{\includegraphics{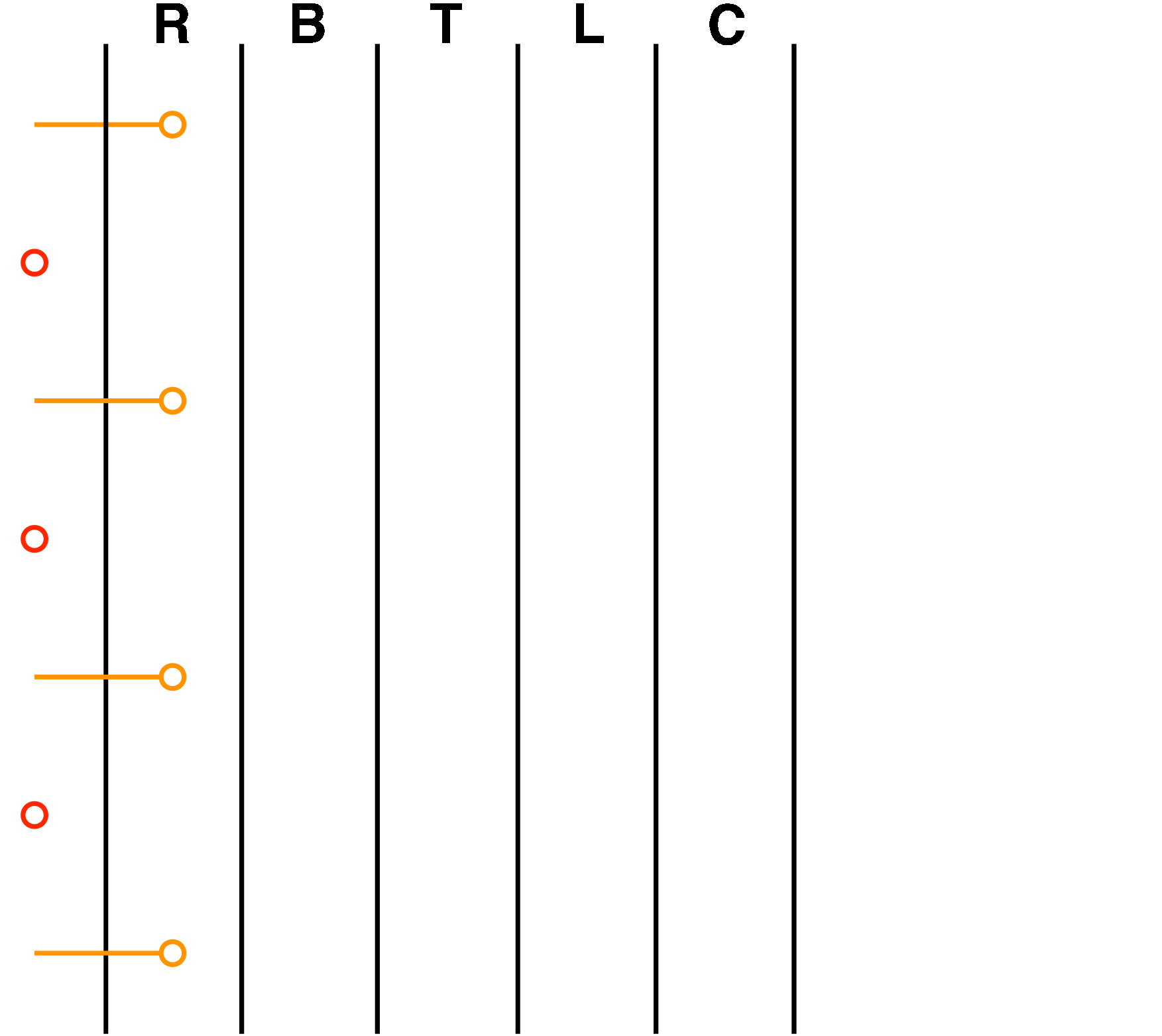}}
\hspace*{3pt}
\resizebox{34mm}{!}{\includegraphics{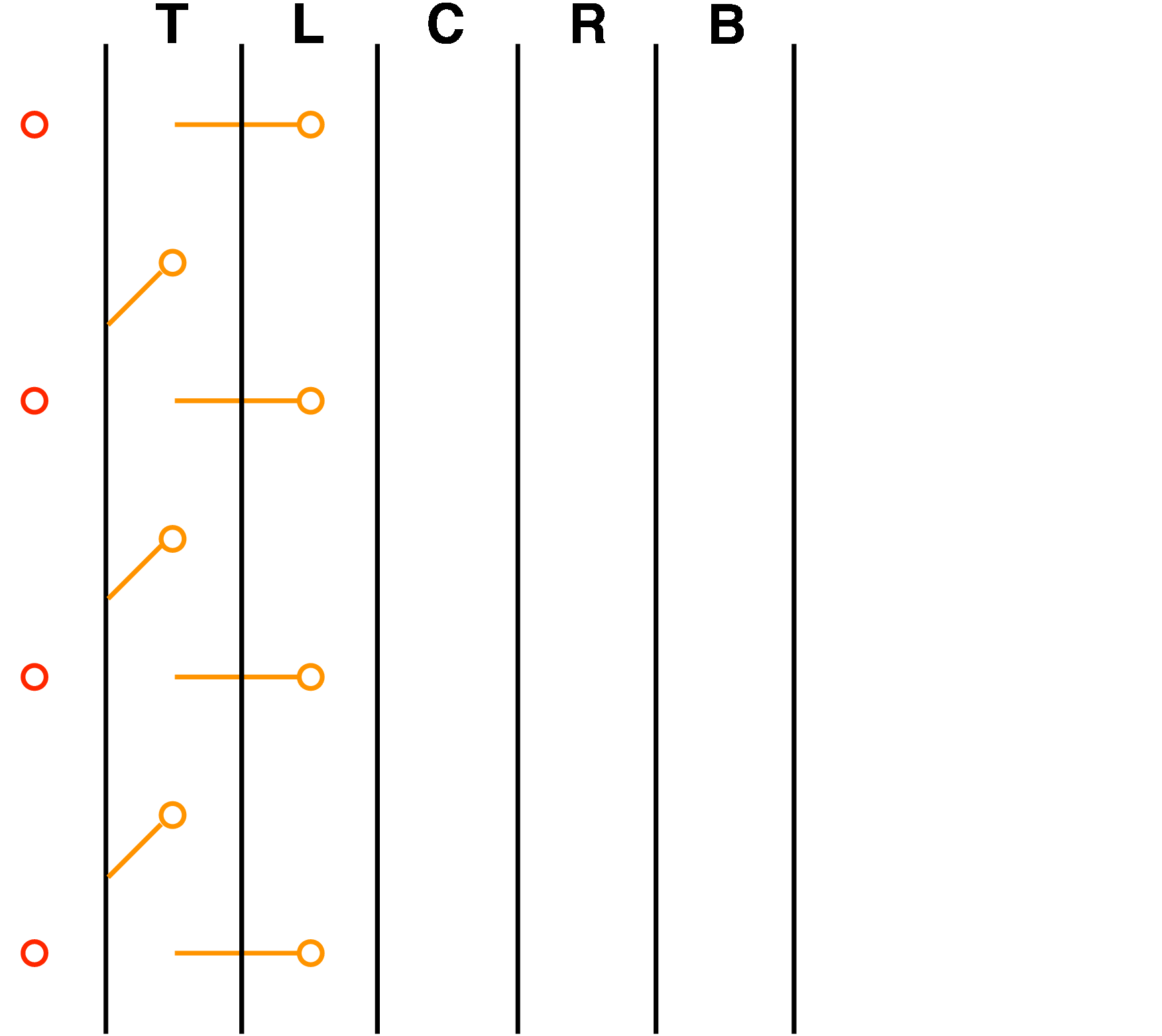}}
\hspace*{3pt}
\resizebox{34mm}{!}{\includegraphics{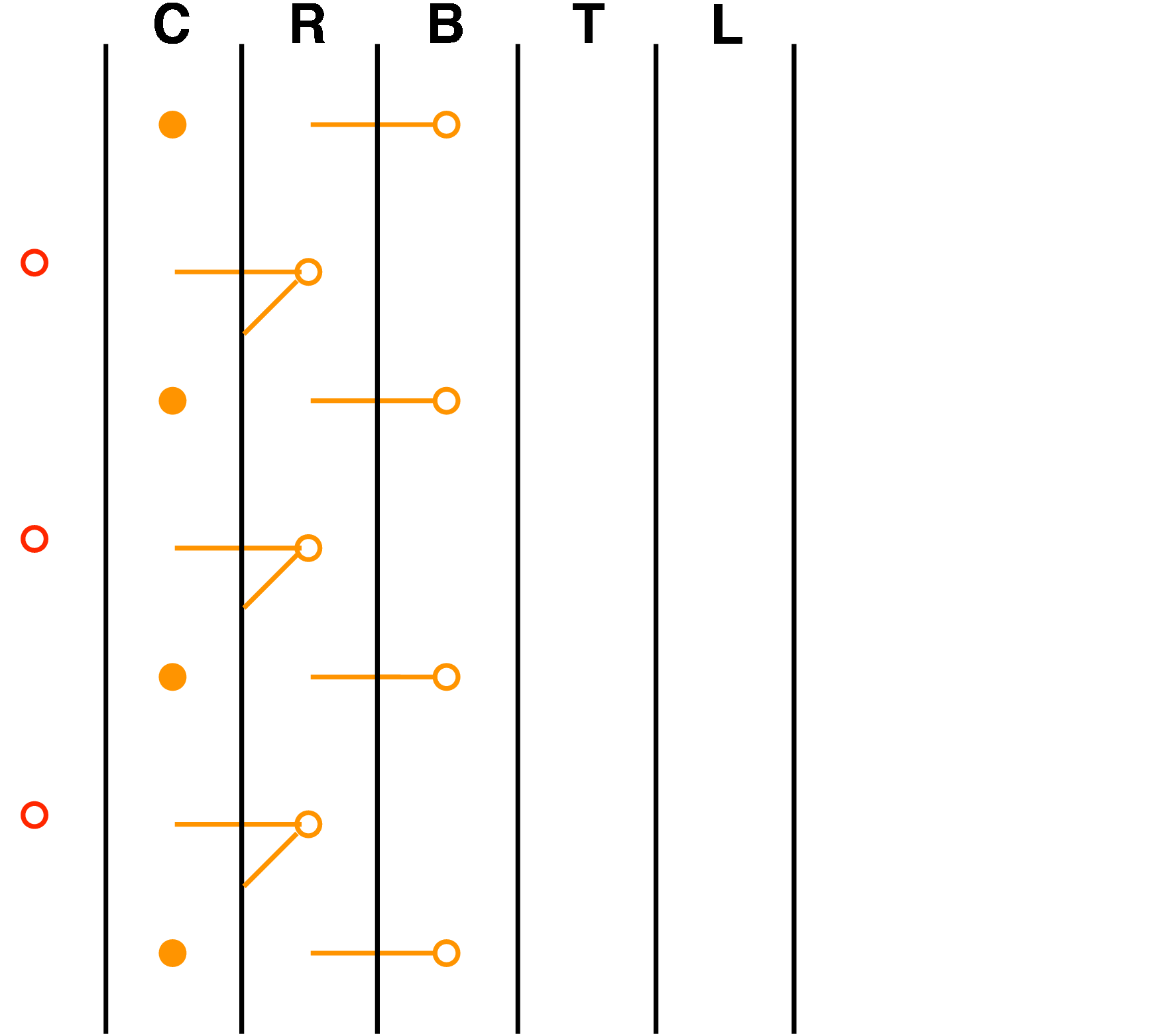}}
\hspace*{3pt}
\resizebox{34mm}{!}{\includegraphics{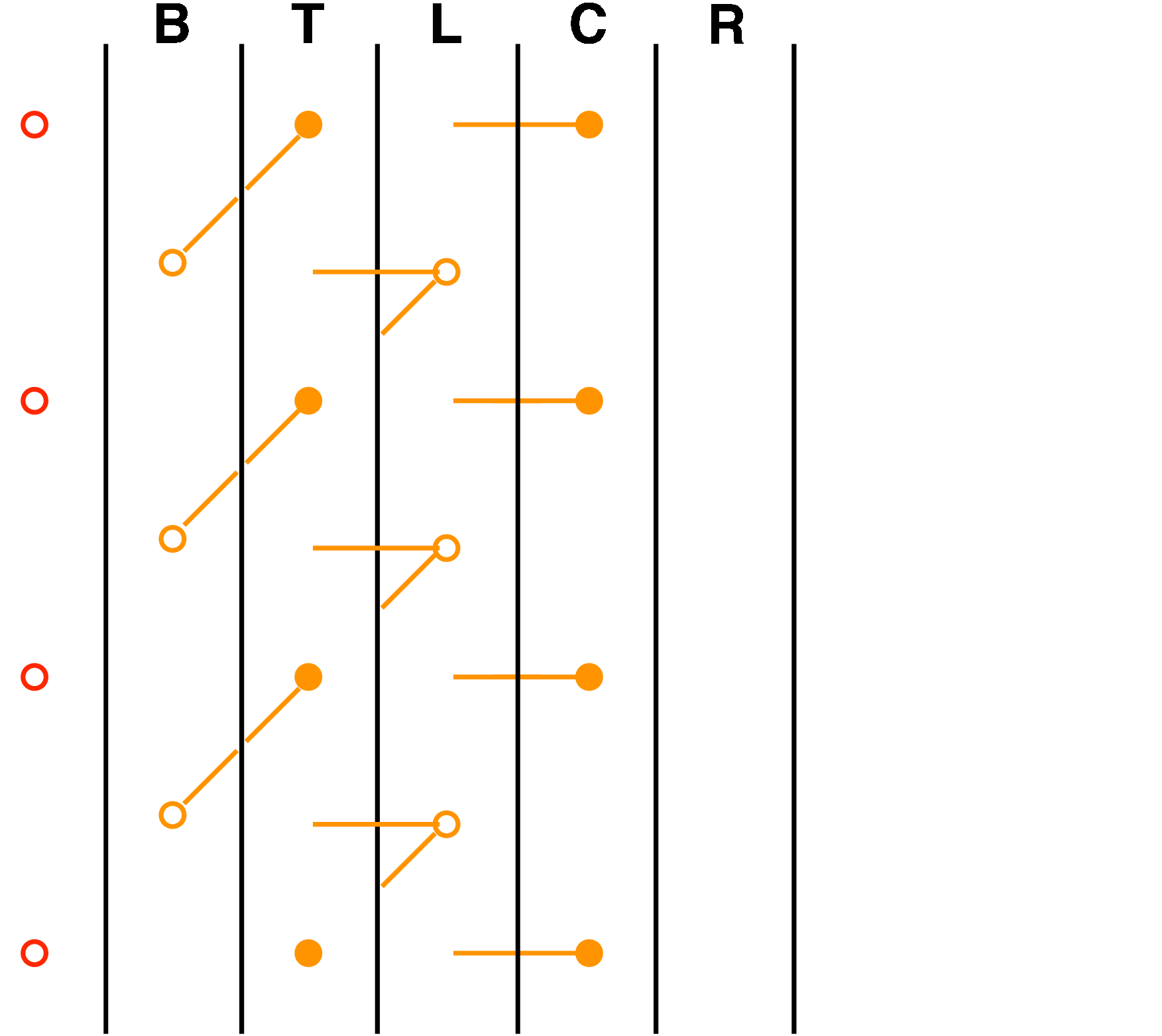}}

\raisebox{-0.5cm}{\resizebox{0.1mm}{!}{\includegraphics{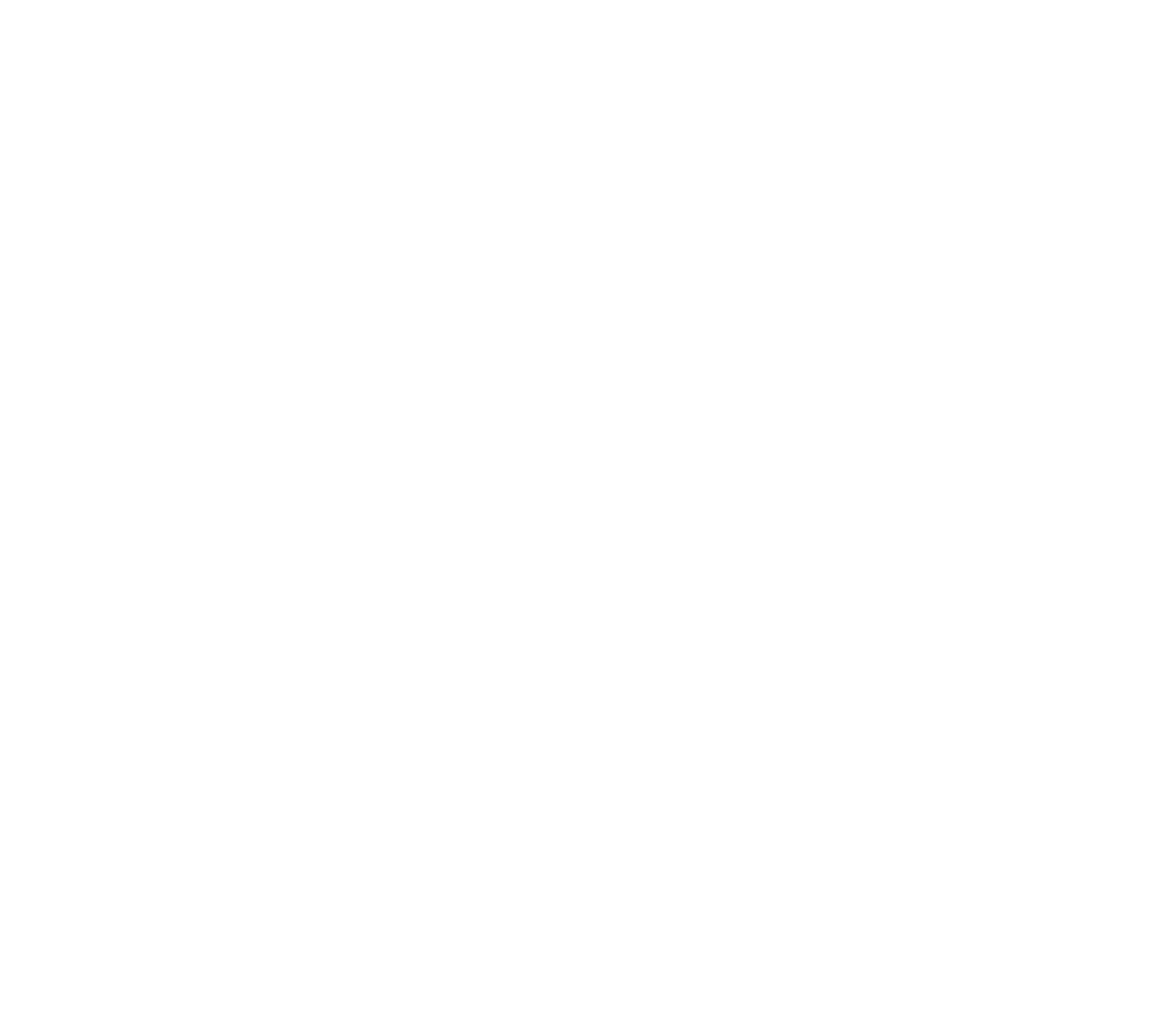}}}

\raisebox{-0.0cm}{\resizebox{34mm}{!}{\includegraphics{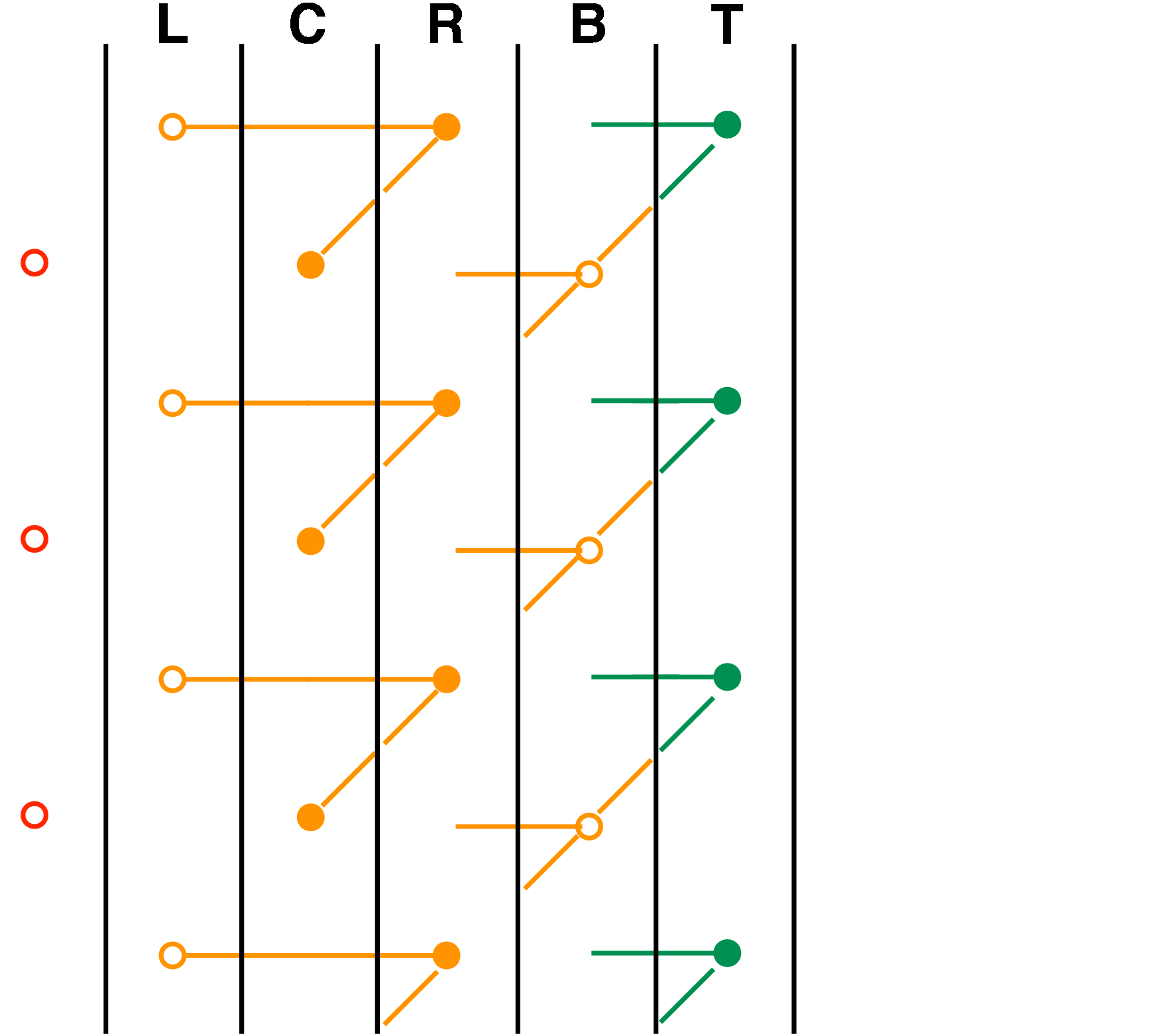}}}
\hspace*{3pt}
\raisebox{-0.0cm}{\resizebox{34mm}{!}{\includegraphics{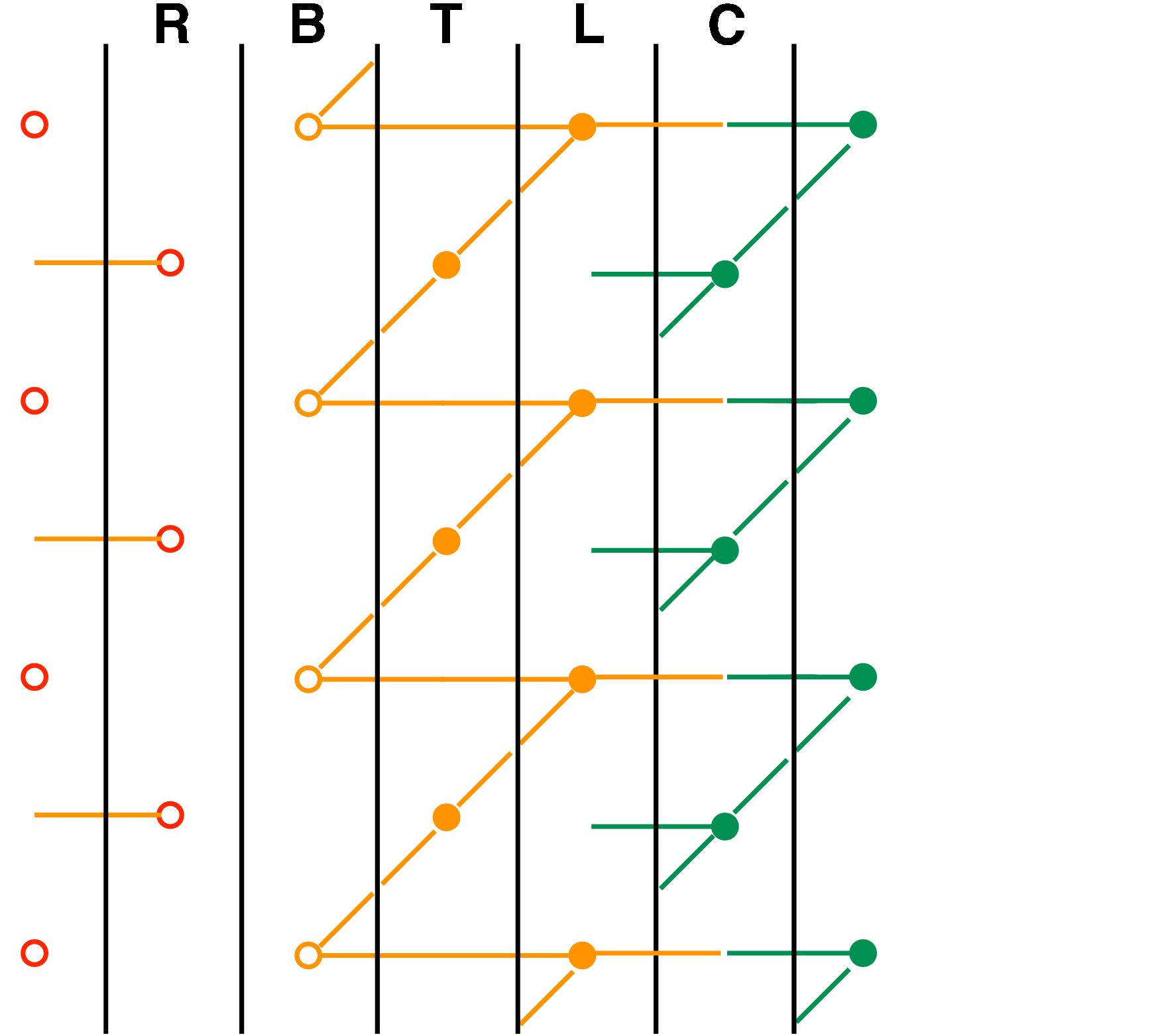}}}
\hspace*{3pt}
\raisebox{-0.0cm}{\resizebox{34mm}{!}{\includegraphics{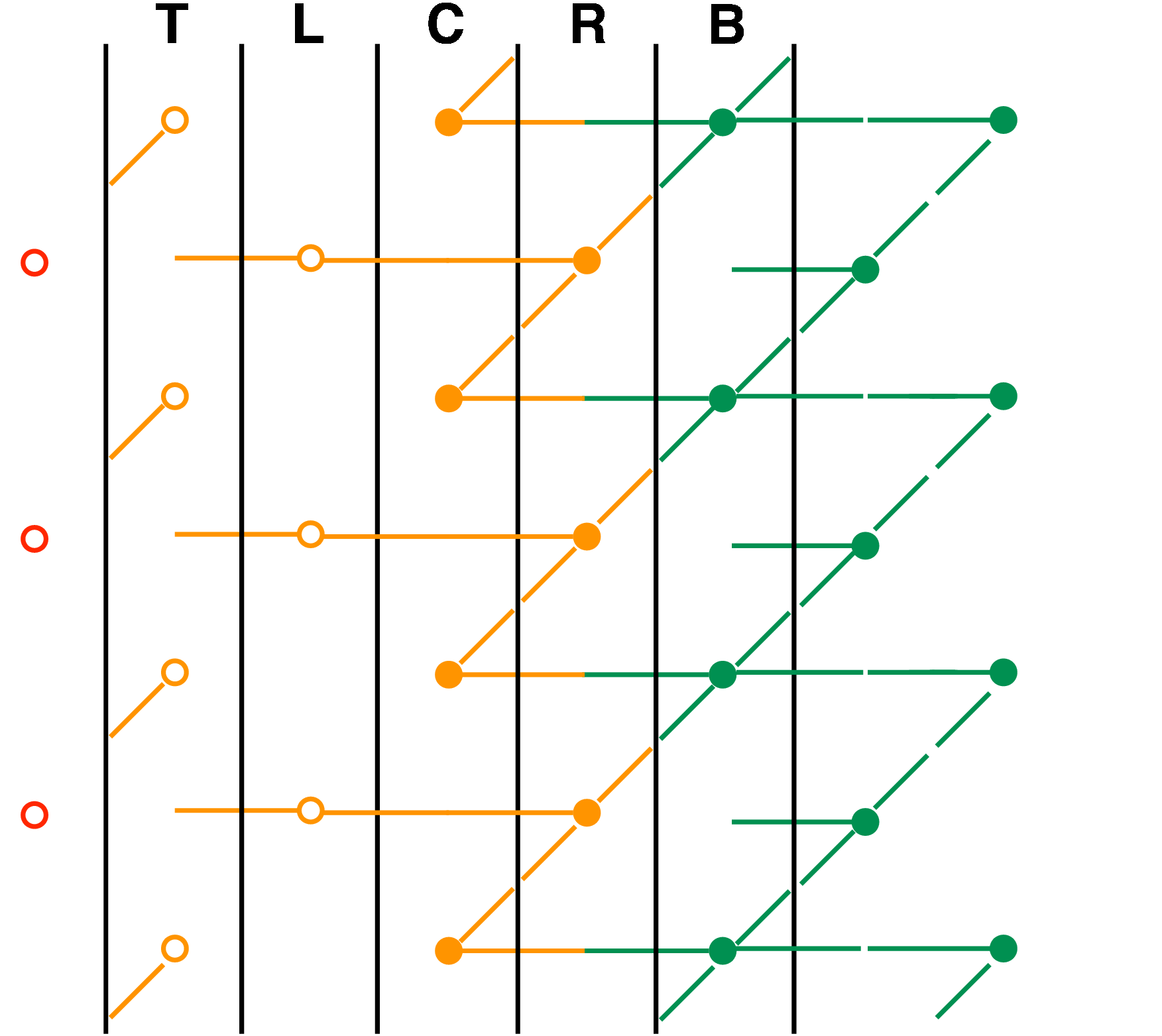}}}
\hspace*{3pt}
\raisebox{-0.0cm}{\resizebox{34mm}{!}{\includegraphics{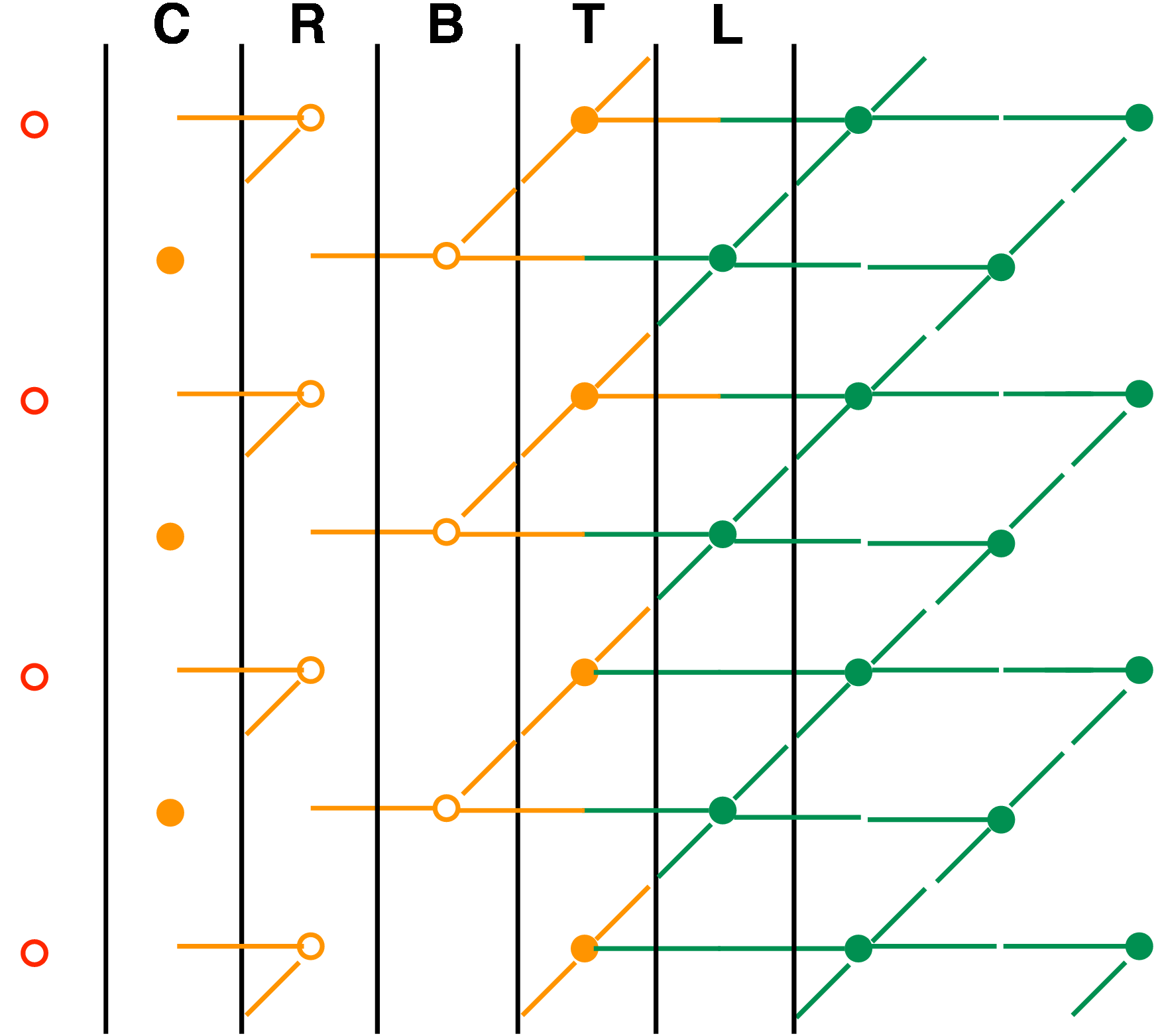}}}

\raisebox{-0.0cm}{\resizebox{0.1mm}{!}{\includegraphics{whitespace.eps}}}
\end{center}
\vspace*{-10pt}
\caption{(Color online) Continuous preparation of a two-dimensional cluster state by the asynchronous network. As described in the caption of Tab.~\ref{table:routing2}, the label over each column indicates how photons are switched in the photonic chips at that point in time. Red circles indicate photons that have not yet entered any modules and so are unentangled. Orange circles indicate photons that have passed through at least one module but not the entire network. Filled orange circles represent photons that have been used as the center photon in a stabilizer measurement. Lines indicate the creation of entanglement between photons. Green lines and green photons indicate cluster photons exciting the network and their complete stabilizers.}
\label{figure:prep3}
\end{figure*}

As in the synchronous network, each photonic chip in the asynchronous network is required to be a three input, three output device, however it is never required to switch in or out three photons simultaneously. Instead (due to the staggering of the input photons) there is only ever a single photon at the middle input or single photons simultaneously at both the top and bottom inputs. It is necessary that in both of these cases individual photons can be selectively routed to or past the photonic module. The six switching settings that are required are illustrated in Fig.~\ref{figure:chip2req} and the switch and chip design in Fig.~\ref{figure:chip2} (and also Fig.~\ref{figure:module1}). Photons are routed through the network according to Tab.~\ref{table:routing2}. Note that photons are never switched to the photonic module in both the A and B chips of a single column. This is so that while the atomic qubits in the A photonic chips are being measured photons are interacting with the atomic qubits in the B photonic chips, and vice versa ($5\delta_t$ is allocated for measurement of the atomic qubit). Because there is no need to delay photons while atomic qubits are measured, the rate of preparation of the cluster is potentially much faster than it is using the synchronous network, $2\delta_t$ per column for a cluster of any size. The overall action of the asynchronous network in preparing a two-dimensional cluster state is illustrated in Fig.~\ref{figure:prep3}.

A comparison of the analogous Figs.~\ref{figure:prep2} and \ref{figure:prep3} shows that the overall action of the synchronous network and the asynchronous network is similar - that is, both networks prepare from unentangled photons a two-dimensional cluster state of any size at some constant rate. However we emphasize several fundamental differences between the two networks: the asynchronous network removes the need for buffering and so the photonic chip required for this network is expected to be simpler to fabricate and to characterize than the chip for the synchronous network. Also, the asynchronous network is potentially able to produce a cluster at a faster rate than the synchronous network. These gains are achieved by increasing the number of chips per row of the cluster from one to five. As it is anticipated that both networks will be constructed by mass fabrication of individual chips which can be individually characterized (so that defective chips can be discarded), it is reasonable to decrease the complexity of each chip at the expense of more chips. 

\section{Consuming the cluster}
\label{EC}
As a final consideration, we examine some of the issues related to the consumption of the cluster to perform computation. In this section we focus on the asynchronous preparation network and for simplicity assume that all optical routing within and between photonic chips is instantaneous. 

As shown in Fig.~\ref{figure:prep3}, the cluster produced by the asynchronous network is a rhombus lattice (where adjacent rows are linked across the diagonal) rather than the square lattice usually associated with a cluster state. This is not an impediment to computing with the cluster. Single qubit gates are performed by consuming part of only a single row of the cluster and so there is no difference between a square lattice and a rhombus lattice.  Performing an interaction between qubits requires consuming correlations between rows \cite{Raussendorf4}. If all measurements required to perform the interaction are completed before either row is used in a non-Clifford gate, no modification to the standard circuits are required. If this condition is not automatically satisfied by the circuit that is being simulated, additional measurements must be performed to simulate repeated identity gates on whichever row measurements are completed first.

As each photon is involved in a total of five stabilizer measurements during preparation of the cluster, there will be a delay of $5\delta_t$ between when photons are initially injected into the network and when the first photons are available for measurement. Once the system has started to prepare the cluster qubits, the delay between photons on each row will be $2\delta_t$, where $\delta_t$ depends strongly on the atom-cavity system used to implement the photonic module. Several systems could be considered, such as Cs, Rb, and NV$^-$, with expected cavity interaction times of 300ns \cite{Boozer1}, 30ns \cite{Trupke1}, and 1ns \cite{Song1} respectively. If we are to assume the slowest of these systems, we are required to measure a single photon approximately every 600ns. 

Detection is required in the $\{\ket{H},\ket{V}\}$ basis, which can be achieved with a polarizing beam splitter, two single photon detectors, and the ability to perform single photon rotations based on measurement results (which can be done via switchable wave plates controlled by the results of detector clicks).  Recent results~\cite{Prevedel1,Bohi1} have demonstrated single photon detection and feedforward on a timescale of 150ns at greater than 99\% fidelity, well below the 600ns required. However, if a longer temporal window is required for measurement, we are able to vary the rate of preparation of the cluster state. Slowing down the cluster preparation network is achieved by increasing the temporal interval between each photon pulse and synchronizing the Stark shift controls on each Q-switch to the repetition rate of each single photon source. Using this approach, only decoherence of the atomic qubit in the photonic module limits how slowly the cluster can be prepared.

An additional benefit to a variable preparation rate is the ability to accommodate the time required to measure the atomic qubit to effect each stabilizer measurement. Assuming the interaction in each photonic module is the light shift method summarized in Section~\ref{TPM}, optical pumping and photoluminescence will most probably be employed to measure the atomic qubit. The asynchronous preparation network as described includes a 5$\delta_t$ window for measurement, but this window can be extended as required by slowing down the cluster preparation network as above. Finally, because each module is only performing a stabilizer measurement involving at most five photons, no matter how large the desired cluster state is, the coherence time of the atomic system only needs to be long enough for five photons to pass through the module between its initialization and measurement.
 
\section{Conclusions and further work}
We have detailed an optical network for the deterministic preparation of arbitrarily large two-dimensional cluster states. Each photonic chip can be independently constructed and characterized before it is incorporated into the network to ensure defect tolerance. Importantly, the cluster state is generated continuously, conditional routing is not required, the size of the preparation network does not depend on the length of the computation, and photon storage is not required if the detector network is placed immediately after the preparation network.

We have presented three distinct versions of the preparation network. The first network can theoretically prepare an arbitrarily large cluster in constant time, but is presented for conceptual reasons only as the practical applicability of this scheme is doubtful. The second is a synchronous network which prepares a cluster, column by column, assuming a photonic chip that incorporates slow-light buffering. The third is an asynchronous network which requires no buffering. This is arguably the optimal technique for cluster preparation in a large scale optical quantum computer. 

As further work, the internal control and construction of the photonic module should be modeled to understand the effect of decoherence and systemic imprecision in the components of the network on the output cluster state. Also, we note that a similar network of photonic modules could be designed to continuously prepare the three-dimensional state required for topological cluster state quantum computing with high threshold \cite{Raussendorf2, Raussendorf3, Fowler1, Devitt2}.

\section*{Acknowledgements}  
The authors acknowledge helpful discussions with C. Myers, T. Ralph, C.-H. Su and T. Tilma. SJD, WJM, and KN acknowledge the support of QAP and MEXT. AMS, ZWEE, ADG, and LCLH acknowledge the support of the Australian Research Council (ARC), the US National Security Agency (NSA), and the Army Research Office (ARO) under contract number W911NF-04-1-0290. ADG is the recipient of an Australian Research Council Queen Elizabeth II Fellowship (project number DP0880466). LCLH is the recipient of an Australian Research Council Australian Professorial Fellowship (project number DP0770715).

\bibliographystyle{unsrt}

\end{document}